\documentclass[12pt,preprint]{aastex}

\def\iso#1#2{\hbox{${}^{\rm #2}{\rm #1}$}}

\def\rpro{{\em r}-process}
\def\spro{{\em s}-process}
\def\ncap{{\em n}-capture}

\def\eg{\mbox{\rm e.g.}}
\def\ie{\mbox{\rm i.e.}}

\def\etal{\mbox{et al.}}

\def\beq{\begin{equation}}
\def\eeq{\end{equation}}
\def\beqar{\begin{eqnarray}}
\def\eeqar{\end{eqnarray}}

\def\gtaprx{ \mathrel{  \vcenter{
                        \offinterlineskip \hbox{$>$}
                        \kern 0.3ex \hbox{$\sim$}    } } }
\def\fun#1#2{\lower3.6pt\vbox{\baselineskip0pt\lineskip.9pt
  \ialign{$\mathsurround=0pt#1\hfil##\hfil$\crcr#2\crcr\sim\crcr}}}

\def\first{1$^{\rm st}$}
\def\second{2$^{\rm nd}$}
\def\third{3$^{\rm rd}$}

\def\bd17{\mbox{BD +17 3248}}
\def\cs22{\mbox{CS 22892-052}}

\begin{document}

\title{Explorations of the {\em r}-Processes: Comparisons between 
Calculations and Observations of Low-Metallicity Stars} 

\author{
Karl-Ludwig Kratz,\altaffilmark{1,2}
Khalil Farouqi,\altaffilmark{2}
Bernd Pfeiffer,\altaffilmark{2}
James W. Truran,\altaffilmark{3} \\
Christopher Sneden,\altaffilmark{4}
and John J. Cowan\altaffilmark{5}
}
\altaffiltext{1}{
Max-Planck-Institut f\"ur Chemie, Otto-Hahn-Institut, 
Joh.-J. Becherweg 27, D-55128 Mainz, Germany}

\altaffiltext{2}{HGF Virtuelles Institut f\"ur 
Struktur der Kerne und Nukleare Astrophysik 
(VISTARS), D-55128 Mainz, Germany;
klkratz, farouqi, bernd.pfeiffer@uni-mainz.de}

\altaffiltext{3}{Department of Astronomy and Astrophysics and 
Enrico Fermi Institute,
University of Chicago, Chicago, IL 60637 and
Argonne National Laboratory, Argonne, IL 60439, USA; truran@nova.uchicago.edu}

\altaffiltext{4}{Department of Astronomy and McDonald Observatory, 
University of Texas, Austin, TX 78712, USA; chris@verdi.as.utexas.edu}

\altaffiltext{5}{Homer L. Dodge Department of Physics and Astronomy, 
University of Oklahoma,
Norman, OK 73019, USA; cowan@nhn.ou.edu}

\begin{abstract}

Abundances of heavier elements (barium and beyond) in many
neutron-capture-element-rich halo stars accurately replicate the solar system
$r$-process pattern.
However, abundances of lighter neutron-capture elements in these stars
are not consistent with the solar system pattern. These comparisons
suggest contributions from two distinct types of
$r$-process  synthesis events -- a so called main $r$-process for the elements
above the \second\  $r$-process peak and a weak $r$-process for the
lighter neutron-capture elements.
We have performed $r$-process theoretical predictions to further explore
the implications of the solar and stellar observations.
We find that the isotopic composition of barium and the elemental Ba/Eu
abundance ratios in $r$-process-rich low metallicity stars can only be
matched by computations in which the neutron densities are in the
range 
23~$\lesssim$~log~$n_n$~$\lesssim$~28, values typical of the
main $r$-process.
For $r$-process  conditions that successfully generate the heavy element
pattern extending down to 
A~=~135, the relative abundance of \iso{I}{129}
produced in this mass region appears to be at least $\sim$ 90\% of the
observed solar value.
Finally, in the neutron number density ranges required for production of the
observed solar/stellar \third\  $r$-process-peak (A~$\approx$~200),
the predicted abundances of inter-peak element hafnium (Z=72,
A~$\approx$~177-180) follow closely those of \third-peak elements
and lead.
Hf, observable from the ground and close in mass number to the \third\ 
$r$-process-peak elements, might also be utilized as part of a new
nuclear chronometer
pair, Th/Hf,  for stellar age determinations.

\end{abstract}



\section{INTRODUCTION}


The nature of rapid neutron-capture nucleosynthesis 
(the \rpro) and its contributions to the abundances of post 
iron-peak elements (Z~$>$~30) were first delineated in pioneering studies 
by Cameron (1957)\nocite{cam57} and Burbidge \etal\ (1957).\nocite{bur57}
The details, however, still remain to be worked out (\eg, see Truran 
\etal\ 2002\nocite{tru02}; Sneden \& Cowan 2003\nocite{sne03a}; 
Cowan \& Thielemann 2004\nocite{cow04}; Cowan \& Sneden 2006\nocite{cow06} 
for recent reviews and discussion). 
The physics of the \rpro\ involves nuclear masses, $\beta$-decay and 
neutron-capture (\ncap) rates, and fission properties of unstable nuclear 
species far from the region of $\beta$-stability.
Laboratory conditions needed to measure properties of such exotic
nuclei are very difficult to produce, but much progress has
occurred over the past decade.
About 50 $\beta$-decay half-lives and 10 nuclear masses have now been
measured for  the lighter (A~$\leq$~140) isotopes in the \rpro\  
production path at neutron freezeout 
(\eg, Kratz \etal\ 2000, 2005a,b\nocite{kra00,kra05a,kra05b}; 
Pfeiffer \etal\ 2001a, 2002\nocite{pfe01,pfe02};
M{\"o}ller \etal\ 2003\nocite{mol03}; Kratz 2006\nocite{kra06}).
Experimental data for heavier \rpro\ isotopes are not yet available.

Additionally, we do not clearly understand the characteristics of the 
stellar or supernova environments in which \rpro\ synthesis occurs.
Circumstantial evidence for the synthesis of the heavy (A~$\ge$~130) \rpro\ 
nuclei in some site associated with massive stars -- with lifetimes 
(production timescales) $\tau_{r-process}$~$\leq$~10$^8$ years -- 
seems compelling.
Proposed sites include both an \rpro\ in a high entropy (neutrino driven)
wind from a Type~II supernova (Woosley \etal\ 1994\nocite{woo94}; 
Takahashi, Witti, \& Janka 1994\nocite{tak94}) and one occurring in the 
decompressed ejecta of neutron star mergers (Lattimer \etal\ 
1977\nocite{lat77}; Rosswog \etal\ 1999\nocite{ros99}; Freiburghaus \etal\ 
1999a\nocite{fre99a}). 
Better understood is the astrophysical site for the slow $n$-captures
that synthesize the heaviest \spro\ nuclei (the so-called
``main'' component): He-fusion zones of low and intermediate mass 
asymptotic giant branch stars (see, \eg, Busso, Gallino, \& Wasserburg 
1999\nocite{bus99}) of significantly longer lifetimes, 
$\tau_{s-process}$~$\geq$~10$^9$ years. 
This difference in the timescales for heavy element enrichment makes
the lowest metallicity (=oldest?) halo stars attractive laboratories for 
empirical insights into \rpro\ synthesis (Truran \etal\ 2002\nocite{tru02}; 
Sneden \& Cowan 2003\nocite{sne03a}; Cowan \& Sneden 2006\nocite{cow06}).

Significant numbers of very metal-poor \rpro-rich halo giants have been 
discovered in the past several decades and analyzed with ever-increasing 
detail and accuracy.
We define these stars to be Galactic halo members that have
[Fe/H]~$\lesssim$~--2,\footnote{
We adopt the usual spectroscopic notations that
[A/B]~$\equiv$ log$_{\rm 10}$(N$_{\rm A}$/N$_{\rm B}$)$_{\rm star}$~--
log$_{\rm 10}$(N$_{\rm A}$/N$_{\rm B}$)$_{\odot}$, and
that log~$\epsilon$(A)~$\equiv$ log$_{\rm 10}$(N$_{\rm A}$/N$_{\rm H}$)~+~12.0,
for elements A and B.
Also, metallicity will be assumed here to be equivalent to the stellar
[Fe/H] value.}
\ncap\ relative overabundance factors of ten or more 
(\eg, [Eu/Fe]~$\gtrsim$~+1), and clear evidence for \rpro\ dominance 
over the \spro\ (usually indicated by [Ba/Eu]~$\lesssim$~--0.7).
Comprehensive abundance analyses involving 10-40 \ncap\ elements
have been published for individual \rpro-rich stars by Westin \etal\ 
(2000; HD~115444\nocite{wes00}), Cowan \etal\ (2002; \bd17\nocite{cow02}), 
Hill \etal\ (2002; CS~31082-001\nocite{hil02}), 
Sneden \etal\ (2003; CS~22892-052\nocite{sne03}), 
Christlieb \etal\ (2004; CS~29497-004\nocite{chr04}) 
and Ivans \etal\ (2006; HD~221170\nocite{iva06}).

There also have been recent larger-sample studies of low metallicity 
\ncap-rich giants.
Johnson \& Bolte (2001)\nocite{joh01} analyzed a set of 22 stars but with 
generally fewer \ncap\ elements per star:  more than 10 elements were detected 
in four stars, and more than five elements in an additional 13 stars.
Employing typically 18 elements per star, Honda \etal\ (2004) have made 
fresh analyses of five of the stars studied by previous investigators, and 
have added two additional ones.
Barklem \etal\ (2005)\nocite{bar05} have presented initial results of a 
halo-star survey specifically designed to identify and supply initial
high-resolution abundance analyses of up to nine elements in new 
\ncap-rich candidates.
Over 40 new \rpro-rich stars were reported in that paper.  
More such stars will undoubtedly be discovered as the high-resolution 
survey work continues.

In all \rpro-rich stars studied to date, the abundance distributions 
for the heavier \ncap\ elements (Z~$\geq$~56, Ba and beyond) 
are nearly identical.
We illustrate this apparently ``standard'' distribution in 
Figure~\ref{f1} for five of these stars.
The small star-to-star scatter in observed abundances about the mean
pattern is dominated by observational/analytical uncertainties.
This uniformity was probably not predictable a priori, since
\ncap\ abundances in \rpro-rich low metallicity stars reflect 
contributions from a single supernova or a small number of supernova 
events, while the solar system (hereafter, SS) abundances result from 
many generations of supernovae.
This suggests that there is a {\it unique} astrophysical site that 
dominates the nucleosynthesis of the heaviest \rpro\ isotopes.
The \rpro\ mechanism for the synthesis of the A $\gtrsim$ 130
isotopes (hereafter, the ``main'' component) must be extremely robust. 

However, abundance patterns in the mass regime below A~$\approx$~130 
for the solar system and very metal-poor stars do not support
such a simple story.
Considering data from carbonaceous chondrite meteorites, 
Wasserburg, Busso, \& Gallino (1996)\nocite{was96} first proposed that 
abundances of the short-lived \iso{Pd}{107} and \iso{I}{129} isotopes in 
primordial SS matter are inconsistent with their having been 
formed in uniform production together with the heavy \rpro\ radioactivities 
(specifically, \iso{Hf}{182}, \iso{U}{235}, \iso{U}{238} and \iso{Th}{232}).
Wasserburg \etal\ argued that these two different mass ranges of 
nuclei require different timescales for production, and thus require
two distinct \rpro\ sites. 
Qian \& Wasserburg (2000)\nocite{qia00} further quantified this idea, 
distinguishing the two \rpro\ occurrences as: (1) H~(high frequency) events 
that are the main source of heavy \rpro\ nuclei (A~$>$~130) but not 
\iso{I}{129}; and (2) L~(low frequency) events that are largely responsible
for light \rpro\ nuclei (A$<$130) including \iso{I}{129}. 
These suggestions formed the initial impetus for distinguishing so-called
``weak" and ``main" \rpro\ components. 
We caution that this scenario is not universally accepted.
For example, a pure \rpro\ origin for the important \iso{Hf}{182} abundance 
has been questioned by Meyer \& Clayton (2000).\nocite{mey00} 
Instead, those authors argue that the enhanced abundance of \iso{Hf}{182} 
with respect to \iso{Pd}{107} and \iso{I}{129} in meteorites is due to the 
injection into the solar nebula of \spro\ material from
outside the edge of the helium-exhausted core of a massive star.
If this interpretation is valid, then one observational basis for two 
\rpro\ sites may not be solid.

Inconsistencies between abundances of lighter and heavier \ncap\
abundances are also found in \rpro-rich low metallicity stars.
If the solar system \rpro\ abundance curve is scaled to fit the
metal-poor-star abundances of elements with Z~$\geq$~56, then 
several elements in the range 39~$\leq$~Z~$\leq$~50 (\eg, Y, Ag) appear 
to have significant under-abundances in the stars.
This result has been found in many of the studies cited above, and
is illustrated in detail in Figures~7 and 8 of Sneden \etal\ (2003).
Additionally, Johnson (2002)\nocite{joh02}, Aoki \etal\ (2005)\nocite{aok05}, 
and Barklem \etal\ (2005)\nocite{bar05} have investigated the 
relationship between the abundances of the light Sr-Y-Zr group and 
the heavier main-component Ba in \rpro-dominated stars, finding a very 
large scatter at lower metallicities.
Travaglio \etal\ (2004)\nocite{tra04} have explored the production of 
Sr-Y-Zr in relation to the heavier \ncap\ elements over a large stellar 
metallicity range.
These results all provide further evidence for a second distinct (``weak'') 
\rpro\ site for the synthesis of isotopes below A~$\approx$~130.   

In this paper we report updated theoretical \rpro\ computations employing 
new nuclear data, and use them to explore several aspects of 
solar system and low-metallicity observational results of \ncap\ elements.
The computations are described in \S2. 
The calculated \rpro\ abundances are compared with selected observational 
data in \S3,
to address: (a) neutron density constraints imposed by barium isotopic and
elemental abundances, (b) the contributions from the two $r$-processes 
to light \ncap\ abundances, and (c) a new connection between 
rare-earth and the heaviest stable elements. 
Finally, in \S4 we discuss the new calculations in terms of 
nuclear chronometers.


\section{{\em r}-PROCESS NUCLEOSYNTHESIS CALCULATIONS}


Our \rpro\ calculations were performed under the conditions of
the waiting-point assumption or
(n,$\gamma$) ${\leftrightharpoons}$ ($\gamma$,n) equilibrium 
(see Kratz \etal\ 1988, 1993, 2005a, 
Kratz 2006\nocite{kra88,kra93,kra05a,kra06}; 
Thielemann \etal\ 1994\nocite{thi94}; 
Pfeiffer \etal\ 1997, 2001a\nocite{pfe97,pfe01}; 
Freiburghaus \etal\ 1999b\nocite{fre99b}).
Although dynamic \rpro\ calculations are feasible today
(see \eg, Freiburghaus \etal\ 1999 and Farouqi \etal\ 2006),
the classical \rpro\ approach can be employed to compare with
the astronomical observations - the goal of this paper.
Even with the simplistic assumptions of constant neutron number density, 
and temperature, and instantaneous freezeout the equilibrium models 
presented here reproduce the solar system abundances well.
The classical approach is also largely independent of a stellar model, 
whereas all recent ``more realistic" calculations imply a specific 
astrophysical environment, \eg, supernovae or neutron-star
mergers, which may require quite different astrophysical parameter sets. 

More recent large-scale, fully dynamic network calculations within 
the SN II high-entropy-wind model have shown (Wanajo \etal\ 
2004\nocite{wan04}; Farouqi \etal\ 2005\nocite{far05}) that there exists 
a rather wide range of correlated astrophysical parameters, 
such as $Y_e$, S, $Y_n$, Y$_{seed}$ and V$_{expansion}$,
within which a robust \rpro\ can be performed. 
However, so far SN models do not yet provide convincingly ``unique'' 
astrophysical conditions within the expected \rpro\ parameter space, 
which would constrain the \rpro. 
In addition, a careful comparison between the classical \rpro\
(\eg, Kratz, \etal\ 1993\nocite{kra93}) with supernova thermodynamic 
trajectories given by Takahashi, Witti \& Janka (1994)\nocite{tak94},
has shown very good agreement (for T$_9$ -- $n_n$) between the 
``waiting-point'' equilibrium and dynamic conditions. 
The assumption of constant $n_n$ and temperature is also a 
good approximation up until the freezeout, as shown by \eg,
Freiburghaus \etal\ (1999b)\nocite{fre99b} and Farouqi \etal\ 
(2006).\nocite{far06}
We also note that the neutron freezeout, which occurs on a very rapid 
timescale (Kratz \etal\ 1993\nocite{kra93}), does not affect the abundances 
up to and slightly beyond the \second\
 peak, and does not have a significant effect on the rare
earth or \third\ \rpro-peak regions (see \eg, Rauscher 2004\nocite{rau04} 
and Farouqi \etal\ 2006\nocite{far06}).

Our calculations to reproduce the total isotopic SS
\rpro\ abundance pattern covered a neutron-density range of 
20~$\leq$~log~$n_n$~$\leq$~30  in steps of 0.5 dex, with the neutron 
exposure times $\tau$($n_n$) and weighting functions $\omega$($n_n$) 
for each $n_n$-component as given in Kratz \etal\ (1993)\nocite{kra93} 
and Cowan \etal\ (1999).\nocite{cow99} 
The upper neutron-density limit is constrained by SS abundances. 
Our results indicate that the highest neutron number densities, 
log~$n_n$~$\geq$~30, make little contribution to the overall abundances -- 
a result that is confirmed by more detailed network calculations 
(\eg, Farouqi \etal\ 2006).
More discussion of the astrophysical parameter choices will be given in \S2.2.
Predicted isotopic abundance trends with neutron density are displayed in 
Figure~\ref{f2}.
The general shapes of the curves are the same for all isotopes:
first, a generally sharp rise to a maximum followed by a more gradual
decline -- if log~$n_n$~$<$~20 abundance runs were to be displayed in
Figure~\ref{f2}, then the rising portions of the curves would be more
apparent for the lightest isotopes (A~=~100 and 120).
The details of the curves are dependent on individual nuclear properties
and will not be discussed further.

Of greater interest for this paper are \ncap\ elemental abundances in 
metal-poor stars.
Therefore, as stellar spectroscopy normally can yield only elemental 
abundances, in Figure~\ref{f3} we show some examples of the \rpro\
results summed by element. 
As in Figure~\ref{f2}, predicted abundances are plotted as a function 
of individual neutron number density log~$n_n$.  
Unsurprisingly, these elemental abundance variations are similar to the
isotopic ones, and the same statement about the ``missing'' parts of the
light isotope trends applies to the elemental curves as well.
The data of these figures show that lighter \ncap\ elements are 
produced in bulk at smaller values of log~$n_n$ than are the heavier ones.


\subsection{Nuclear Data Input}


The best agreement with SS \rpro\ abundances is obtained when
we employ: {\it (a)} the nuclear mass predictions from an extended 
Thomas Fermi model with quenched shell effects far from stability 
(\ie, ETFSI-Q; Pearson, Nayak, \& Goriely 1996); and {\it (b)} the 
$\beta$-decay properties from QRPA-calculations for the Gamow-Teller(GT) 
transitions based on the methods described in M\"oller \& Randrup 
(1990\nocite{mol90}; see also M\"oller \etal\ 1997\nocite{mol97});
and the first-forbidden strength contributions from the Gross theory of 
$\beta$-decay (M\"oller, Pfeiffer \& Kratz 2003\nocite{mol03}).

We have previously tested quite a number of global mass models in our
classical \rpro\ calculations (see, \eg, Kratz \etal\  
1998, 2000, 2005a,b\nocite{kra98,kra00,kra05a,kra05b}; 
Pfeiffer \etal\  1997, 2001a\nocite{pfe97,pfe01}; Kratz 2006\nocite{kra06}). 
For our present paper we have chosen to use the ``quenched'' mass formula 
ETFSI-Q for the following reasons.
In principle, a weakening of the strong N=82 shell below \iso{Sn}{132} had
already been suggested by Kratz \etal\ (1993)\nocite{kra93} from, at 
that time, scarce experimental indications in the phase transitional 
region around A=100.
Following the microscopic, self consistent description of ``shell quenching''
far from stability in the spherical mass model 
HFB(Hartree-Fock-Bogoliubov)/SkP(Skyrme force) of Dobaczewski 
\etal\ (1996),\nocite{dob96} Pearson \etal\ (1996)\nocite{pea96}
implemented this behavior in an algebraic procedure for the 
shell closures N=50, 82 and 126 into their ``unquenched'' deformed 
ETFSI mass formula, hereafter called ETFSI-Q.
The big improvement in changing the \rpro\ matter flow at the
closed-shell ``bottle-necks'',  and consequently in reproducing the overall
SS \rpro\ abundance pattern by our classical waiting-point 
calculations, was evident immediately (Kratz \etal\ 1998;
Pfeiffer \etal\ 1997, 2001a).

Today, several other microscopic HFB models have become available with 
the same overall quality as the ETFSI-Q approach in terms of the 
root-mean-square differences between experimentally known and predicted 
masses (see \eg, Lunney \etal\ 2003\nocite{lun03}; Rikovska-Stone
2005\nocite{rik05}; Pearson 2004).\nocite{pea04}
However, these models do not reproduce well the experimental N~=~82 
shell-gap behavior around double-magic \iso{Sn}{132}, where the dominant 
bottle-neck in the total \rpro\ matter flow occurs (Kratz \etal\ 2005a,b;
Kratz 2006, and further discussion below).
Furthermore, the recent HFB models show a rather unphysical chaotic 
behavior in their $S_{2n}$ systematics, indicating that the treatment 
of pairing in these approaches is inadequate (see, \eg, Rikovska-Stone 2005, 
Kratz \etal\ 2005a,b; Kratz 2006).
Summarizing, although ETFSI-Q may not be the ultimate mass model for
\rpro\ calculations, it is preferred for our computations because it best 
reproduces the experimentally measured nuclear-structure quantities of 
more than 20 \rpro\ isotopes in the A~=~130 $r$-abundance peak region.

Our nuclear database has been improved by including recent experimental
results and, based on a better understanding of the underlying shell
structure, new theoretical predictions. 
By now, altogether more than 50 isotopes lying in the \rpro\ path under 
log~$n_n$~= 20 freezeout conditions between \iso{Fe}{68} and \iso{Sb}{139} 
have been measured, at least via their $\beta$-decay half-life. 
Of particular importance are the new spectroscopic data on N~$\simeq$~82 
Ag, Cd, In., Sn, and Sb isotopes, which have led to a better understanding 
of the A~$\simeq$~130 $r$-abundance peak as the major bottle-neck for the 
\rpro\ matter flow to the heavier elements (see, \eg, Kratz \etal\ 
2000, 2005a,b, 2006).\nocite{kra00,kra05a,kra05b,kra06}
In this context, presumably the measurement of utmost importance is the 
determination of the high (energy involved in a $\beta$-decay) 
$Q_{\beta}$ value of N~=~82 \iso{Cd}{130}, 
which represents the isobaric mass difference between \iso{Cd}{130} 
and its daughter \iso{In}{130} (Dillmann \etal\ 2003\nocite{dil03}; 
Kratz \etal\ 2005b).
This experimental value is in clear disagreement with predictions from 
all older ``unquenched'' global mass models (\eg, the GTNM or Gross Theory 
of Nuclear Masses, of Hilf \etal\ 1976\nocite{hil76}; FRDM (Finite Range
Droplet Model) of 
M\"oller \etal\ 1995\nocite{mol95}; ETFSI-1 of Aboussir \etal\ 
1995\nocite{abo95}), as well as from the recent series of more microscopic 
HFB approaches (\eg, Goriely \etal\ 2001, (HFB-2)\nocite{gor01}; 
Samyn \etal\  2004, (HFB-8); Goriely \etal\ (HFB-9)\nocite{gor01}, 2005; 
Rikovska-Stone 2005).
The experimental value is only in agreement with the ``quenched'' mass 
models HFB/SkP (Dobaczewski \etal\ 1996\nocite{dob96}), ETFSI-Q 
(Pearson \etal\ 1996\nocite{pea96}) and the latest nuclear-mass evaluation 
of Audi \etal\ (2003).\nocite{aud03}

An analysis of the discrepancies between measured and calculated 
$\beta$-decay properties reveals considerable improvement over the 
earlier evaluation. 
For potential \rpro\ progenitor isotopes with half-life 
T$_{1/2}$~$\leq$~0.2~s, the total error is now a factor of two; for 
the $\beta$-delayed neutron emission probability, $P_n$~$\geq$~1~$\%$,
the mean deviation
is a factor of three. 
Moreover, we have performed a careful parameter study of the 
``robustness'' of T$_{1/2}$ and  $P_n$  predictions of the N~$\simeq$~126 
waiting-point nuclei forming the A~$\simeq$~195 \rpro-peak, where no
experimental data are available at all.
Again, our model predictions lie within a factor of two of the observed
abundances -- this is in contrast to earlier calculations, often uncertain 
by an order of magnitude or more, that were dominated by uncertain
nuclear physics data for the most neutron-rich nuclei.  
Taken together, this gives us confidence in the reliability of our
nuclear-physics input to the subsequent \rpro\ calculations.


\subsection{Abundance Fits to the Sun and \cs22}


As our calculations demonstrate, a range of neutron densities are 
required to reproduce all major features of the SS meteoritic 
\rpro\ isotopic abundance distribution.
However, we remind the reader that the {\it total} meteoritic abundances are 
generally combinations of both \spro\ and \rpro\ contributions.
Because the \spro\ proceeds along the valley of stability and depends mostly on
neutron capture cross sections that are directly measurable in the lab, its fraction of
the total can be estimated. 
This is done either empirically by fitting a smooth curve to the $N\sigma$ 
versus mass distribution of isotopes that can only be synthesized in the 
\spro\ (K{\"a}ppeler \etal\ 1989\nocite{kap89}; Burris \etal\ 
2000\nocite{bur00}; Simmerer \etal\ 2004\nocite{sim04}; Cowan \etal\ 
2006\nocite{cow06b}), or theoretically by computing an abundance set 
with conditions that correspond to those expected in stellar interior 
He-fusion zones (Arlandini \etal\ 1999\nocite{arl99}; Travaglio \etal\ 
2004\nocite{tra04}).
Then the chosen \spro\ abundance set is subtracted from the total 
SS abundances to yield the \rpro\ set.
The resulting \rpro\ abundances may be very accurate for those isotopes 
with little \spro\ contribution (\eg, \iso{Eu}{151} and \iso{Eu}{153}), 
but have significant uncertainties for isotopes where the \spro\ fraction 
is dominant (\eg, \iso{Ba}{138}, the most abundant of the seven 
naturally-occurring Ba isotopes).
This caution should be kept in mind in all SS \rpro\ abundance 
comparisons, but see also a recent attempt at directly 
predicting the SS \rpro\ abundances for several 
rare-earth elements (Den Hartog \etal\ 2006).\nocite{den06}

Attempts to fit predicted \rpro\ abundances to the SS values
require assumption of a continuous addition of a small number of 
individual neutron density components, with a varying \rpro\ path related 
to contour lines of constant neutron separation energies in the range of 
4--2~MeV (the latter being determined by the combination of 
neutron number density and temperature).
The number of such components needs to be at least four to match the 
relative abundances of the three \rpro\ abundance peaks (A~=~80, 130, 
and 195) and abundances in the actinide region 
(Kratz \etal\ 1993).\nocite{kra93}
Less than 20 components are sufficient to match the detailed \rpro\ 
SS abundances (Pfeiffer \etal\ 1997\nocite{pfe97} and 
Cowan \etal\ 1999\nocite{cow99}).
Adding more components produces no further improvement, given the present
uncertainties in both \rpro\ computations and \rpro\ SS abundances.

While such a procedure is both largely site-independent and mainly 
intended to produce a good fit to solar \rpro\ abundances, it can 
also provide information regarding the conditions that a ``real'' 
\rpro\ site has to fulfill.
The fit is performed by adjusting the weight of the individual components
(or different neutron separation energies $S_n$($n_n$,$T$)) and the time 
duration $\tau$ for which these (constant) conditions are experienced, 
starting with an initial abundance in the Fe-group.\footnote{
We have also performed computations that employ a ``Zr seed'' beyond 
Z~=~40, N~=~50, and the results are similar.}
For a given (arbitrary) temperature, $S_n$
is a function of neutron number density $n_n$.
The addition weights $\omega$($n_n$) and process durations 
$\tau$($n_n)$ have a behavior similar to powers of $n_n$.
This corresponds to a linear relation in log~$n_n$ and is already observed
when taking a minimum of three components in order to fit the \rpro\
abundance peaks (Kratz \etal\ 1993\nocite{kra93}, 
Cowan \etal\ 1999).\nocite{cow99}
This approach (although only a fit and not a realistic site calculation) 
is reasonable to the extent that such a continuous dependence on physical 
conditions can be expected to reflect the range of conditions appropriate 
to the (yet unidentified) astrophysical site.

The effect of the weighting of the sums of the various components on
the abundance fit is shown in the lower panel of Figure~\ref{f4}.
Relatively small weighting factors for the highest neutron number
densities and larger weighting values for lighter neutron density
components yield good fits to the data.
Overall the best fit to this region of the SS abundances (from
A~$\approx$~130 to the actinides) is that of log~$n_n$~=~20 to 26--27.
                                                                                
For a choice of a typical freezeout temperature (in billions of
degrees K) of T$_9$~=~1.35 (see \eg, Cowan \etal\ 1999\nocite{cow99}), 
the necessary addition of components (with \rpro\ paths of a corresponding 
neutron separation energy $S_n$ over a duration $\tau$) can be expressed 
in the form $\omega$($n_n$)~$\simeq$ 8.36$\times$10$^{6}n_n^{-0.247}$ 
and $\tau$($n_n$)~= 6.97$\times$10$^{-2}n_n^{0.062}$~sec.
This algebraic/exponential fit is in good agreement with full network
calculations (Farouqi \etal\ 2006\nocite{far06}) for the high entropy
wind scenario in terms of \rpro\ seed nuclei as a function of
entropy/neutron-density.
When restricting the fit to the mass regions around the \rpro\ peaks
(A $\simeq$ 80, 130, 195), where the paths come closest to stability, model
extrapolations need not be extended far into unknown territory. Even
a significant amount of experimental information is available
(Pfeiffer \etal\ 2002\nocite{pfe02}, Audi \etal\ 2003\nocite{aud03};
Kratz \etal\ 2005a, Kratz 2006\nocite{kra05a,kra06}),
and is  now being  utilized in the calculations.
These power laws in $n_n$ play roles comparable to that of the assumed
exponential addition of neutron exposures in the classical \spro.

The relationship between component abundances computed at single neutron 
densities to their superposition is illustrated in Figure~\ref{f4}. 
In the top panel we show individual, unweighted calculations, ranging from
log~$n_n$~= 24 to 30, superimposed on the solar isotopic abundances 
(black dots).
We have arbitrarily normalized all curves to match the abundances at A = 195.
There are several trends evident in this plot. 
First, larger neutron number densities (log~$n_n$ $\geq$ 25), in general, 
are required to synthesize the heaviest elements (\ie, the 
\third\ peak and actinide regions). 
The larger values of $n_n$ are needed to push the so-called ``\rpro\ path'' 
-- defining where these nuclei are produced -- far enough away from 
stability, and to very neutron-rich regions, so that the synthesized 
radioactive nuclei subsequently decay back to high enough mass numbers  
(\eg, see Figure~1 in Cowan \& Thielemann 2004).\nocite{cow04}
However, the highest neutron density (log~$n_n$ = 30) 
overproduces the heavy region and the Eu region (A $\approx$ 150).
Additionally, the largest neutron densities produce ratios of \third\  
peak to interpeak abundances that are clearly different than those 
observed in the solar system. 
This occurs because such a high value of $n_n$ pushes the \rpro\ path 
too far into the radioactive region and consequently overproduces the 
heaviest nuclei.
On the other hand the lighter neutron number densities
log~$n_n$~= 24 do mimic the A~$\approx$~150 range, but are clearly
inadequate to synthesize the \third\ peak abundances.
This reinforces the idea that multiple neutron exposures are required to 
reproduce the three different SS peaks -- one exposure will not be adequate.

We show in Figure~\ref{f5} the detailed steps in one such typical 
\rpro\ addition calculation.
These calculations include explicitly $\beta$-decays back to pseudo (\ie,
long-half life) stability, but do not include $\alpha$-decays 
from the trans-lead region.
In the top panel of this figure, the summation of five individual 
neutron number density components in the range log~$n_n$~=~20--22 have been 
tuned to reproduce the A~=~80-100 SS isotopic \rpro\ abundances. 
This composition is clearly inadequate to reproduce the abundances of 
the heavier isotopes. 
As shown in the second panel of Figure~\ref{f5}, addition of higher 
neutron number density components (up to log~$n_n$~=~24) mostly reproduces 
the A~=~130 \rpro\ abundance peak. 
Therefore, we estimate that the division between ``weak'' and ``main'' 
\rpro\ density regimes occurs at log~$n_n$~$\simeq$ 23.0~$\pm$~0.5.
Likewise, to form the heaviest stable elements, \third\ \rpro\ peak, 
it is necessary to include neutron number density components at least up 
to log~$n_n$ $\simeq$ 26 (Figure~\ref{f5}, third panel). 
However, even this neutron density is still insufficient to reproduce the 
abundances of the trans-lead region (including Th and U), which requires 
log~$n_n$~$\simeq$~28 as shown in the bottom panel of the figure. 

In Figure~\ref{f6} the illustration is reversed.
Considering our best-fit weighted \rpro\ distribution (the bottom panels
of both Figures~\ref{f5} and \ref{f6}) leads to a few comments.
First, the \rpro\ parts of the heaviest stable elements Pb
(Z~=~82, A~=~206--208) and Bi (Z~=~83, A~=~209) in the SS 
are principally due to 
$\alpha$-decays of nuclei along the radioactive decay chains extending 
through the actinide region. 
Their observed abundances are reproduced well by our calculations.
We show in the subsequent panels (starting at the top with the 
highest values only) how progressively adding additional lower neutron 
number density components leads to a better and better fit of the solar 
system isotopic abundances -- specifically, in the second panel at the 
rare-earth region, the third panel the A~=~195 peak and in the bottom 
panel the complete SS abundance pattern (see Cowan \etal\ 1999).
We note also, as illustrated in Figure~\ref{f5}, that very low neutron 
densities (log~$n_n$~$\simeq$~20) are insufficient to reproduce the first 
\rpro\ peak, and very high densities (log~$n_n$~=~30) overproduce the 
\third\ peak and the actinides, as shown in Figure~\ref{f4}.  

\begin{figure}
\plotone{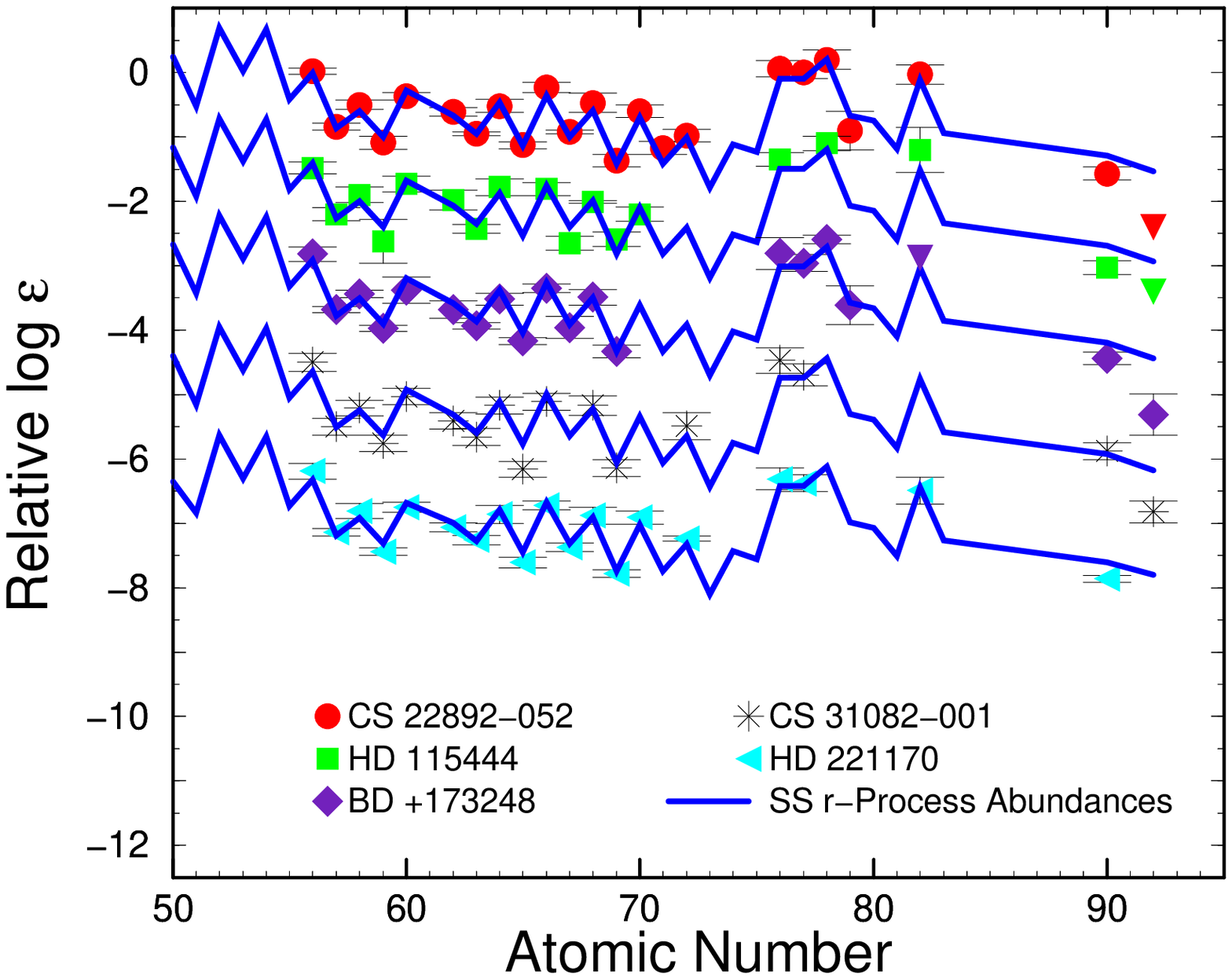}
\caption{Abundances for elements with Z~$\geq$~56 for five \rpro-rich 
Galactic halo stars (Cowan \& Sneden 2006).
Top to bottom the stars are \cs22\ (filled circles, Sneden \etal\ 2003), 
HD~115444 (filled squares, Westin \etal\ 2000), 
\bd17\ (filled diamonds, Cowan \etal\ 2002), 
CS~31082-001 (stars, Hill \etal\ 2002), and HD~221170 (filled 
triangles, Ivans \etal\ 2006).
The vertical scale for the \cs22\ abundance set is true, and abundances 
of all of the stars have been vertically scaled downward for
display purposes. 
Each of these stellar abundance sets is overlaid with the scaled 
SS \rpro\ abundance distribution that (approximately) 
best fits the
observed abundances (solid lines).
\label{f1}}
\end{figure}

\begin{figure}
\epsscale{0.9}
\plotone{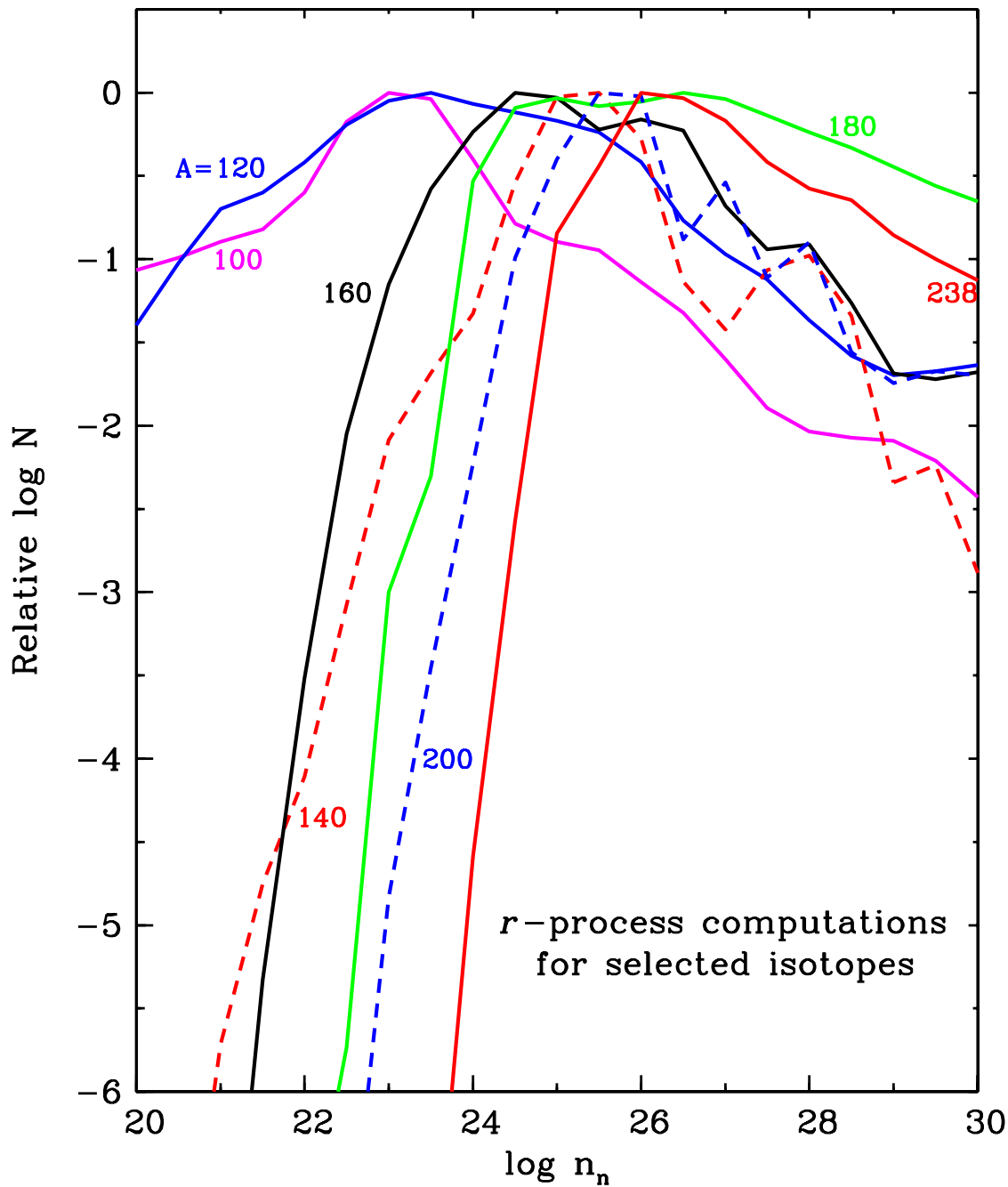}
\caption{
Representative \rpro\ isotopic abundance predictions as a
function of individual neutron number density $n_n$.
The purpose of this figure is to show the relative efficiency of the
\rpro\ in producing various \ncap\ masses as $n_n$ increases.
Therefore for display purposes here, the distribution for each isotope
has been divided by its maximum value, so that all curves peak at
normalized log~N~$\equiv$~0.0,
where N is a number density abundance.   
\label{f2}}
\end{figure}

\begin{figure}
\epsscale{0.9}
\plotone{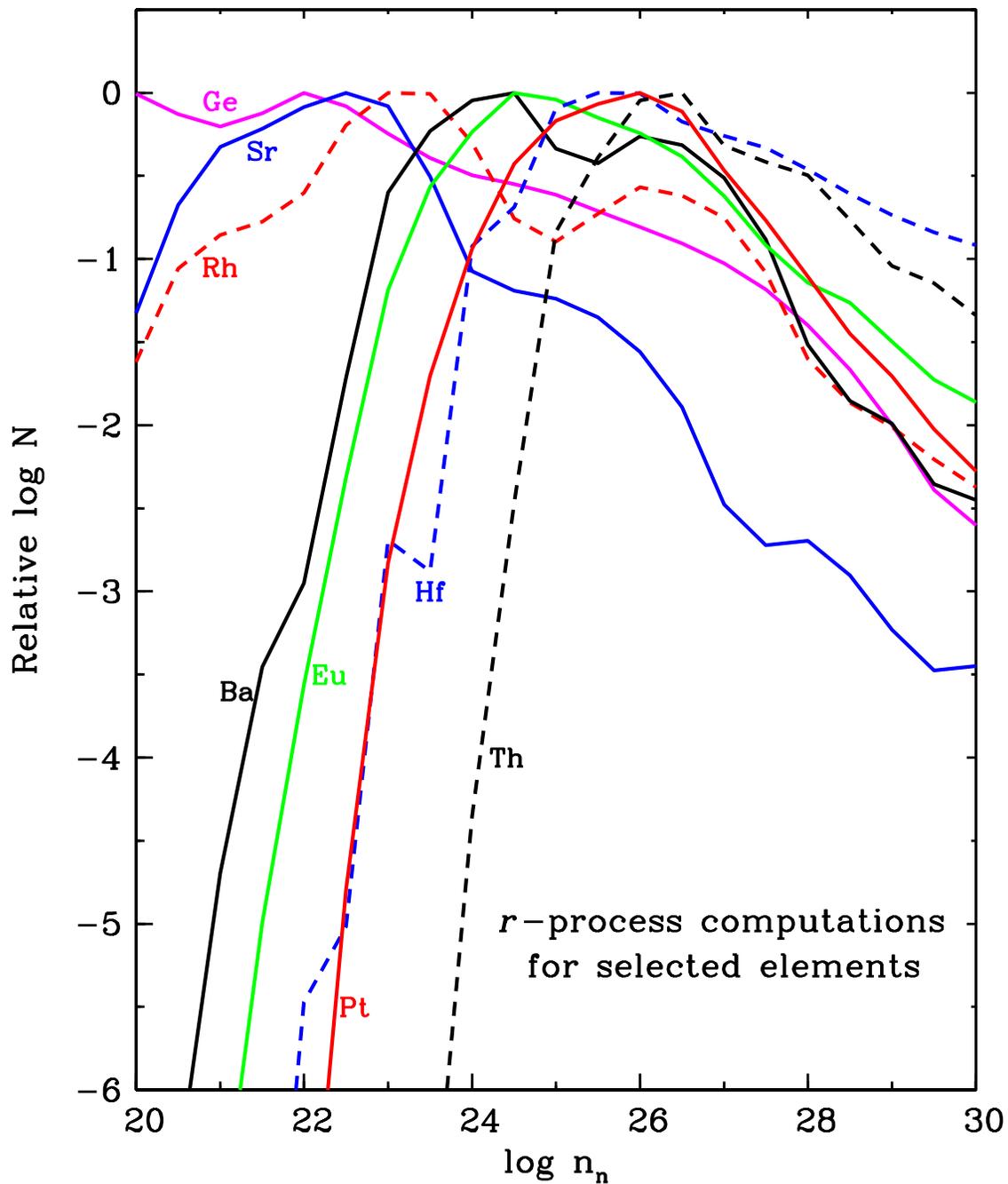}
\caption{
Representative \rpro\ elemental abundance predictions as a 
function of individual neutron number density $n_n$.
The distribution for each element has been divided by its maximum value, 
so that all curves peak at normalized log~N~$\equiv$~0.0.
\label{f3}}
\end{figure}

\begin{figure}
\epsscale{0.9}
\plotone{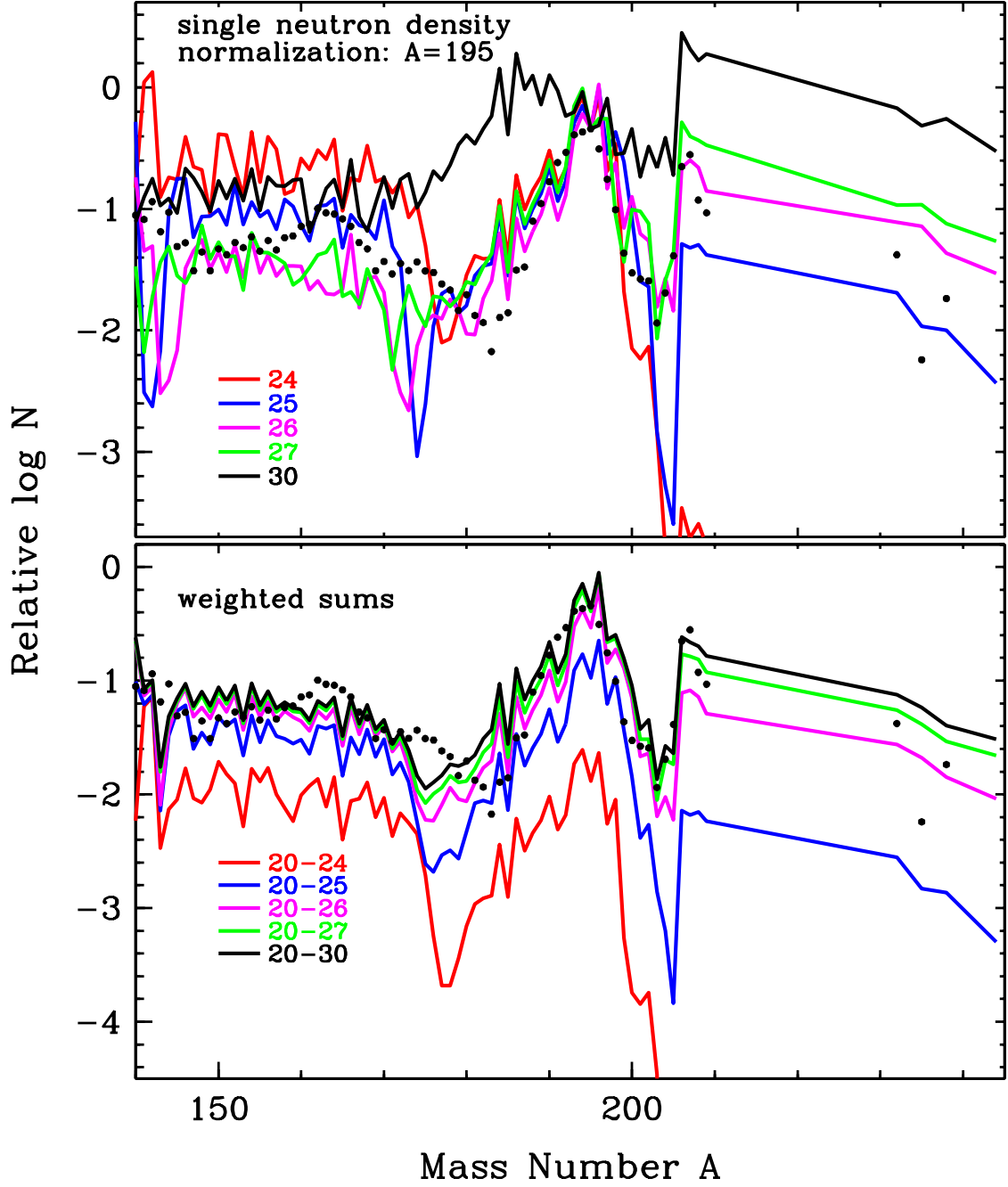}
\caption{
The relationship between component abundances computed at single neutron
densities and their weighted sums.
The top panel shows solid curves representing single neutron densities 
chosen in the range log~$n_n$~= 24 to 30, and solid circles representing 
solar isotopic abundances (black dots).
The curves are shifted to agree with the solar value at A~=~195.
The bottom panel shows the  sums of the individual components 
weighted by the form 
$\omega$($n_n$)~$\simeq$ 8.36$\times$10$^{6}n_n^{-0.247}$.
\label{f4}}
\end{figure}

\begin{figure}
\epsscale{0.70}
\plotone{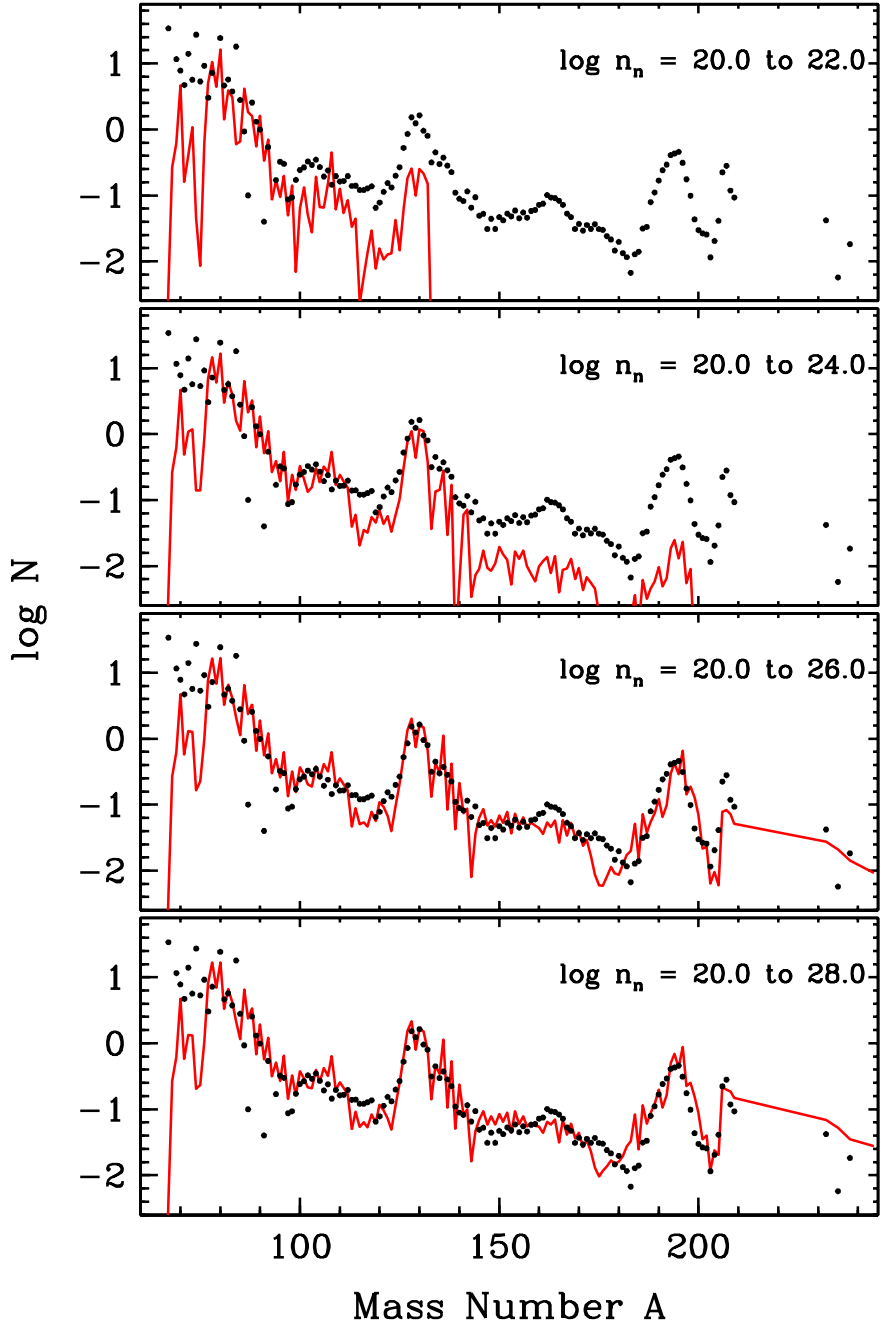}
\caption{
Comparison of meteoritic SS \rpro-only isotopic abundances to 
weighted sums of \rpro\ computations for different neutron densities ranges.
The SS abundances (black points) (K{\"a}ppeler \etal\ 
1989; Wisshak \etal\ 1998; O'Brien \etal\ 2003; 
see listing in Cowan \etal\ 2006) 
are on the standard meteoritic scale in which log N$_{\rm Si}$~=~6.
The four panels show from top to bottom the effect of incorporating 
progressively higher ranges of $n_n$.
The top panel predictions span 20.0~$\leq$~log~$n_n$~$\leq$~22.0, 
adequate only for matching the lightest isotopes.
The bottom three panels successively add more neutron density components
weighted to simultaneously match the greatest mass range of nuclei.
The values displayed here are ones taking into account $\alpha$ and 
$\beta$ decays of nuclei back to stability. 
\label{f5}}
\end{figure}

\begin{figure}
\epsscale{0.8}
\plotone{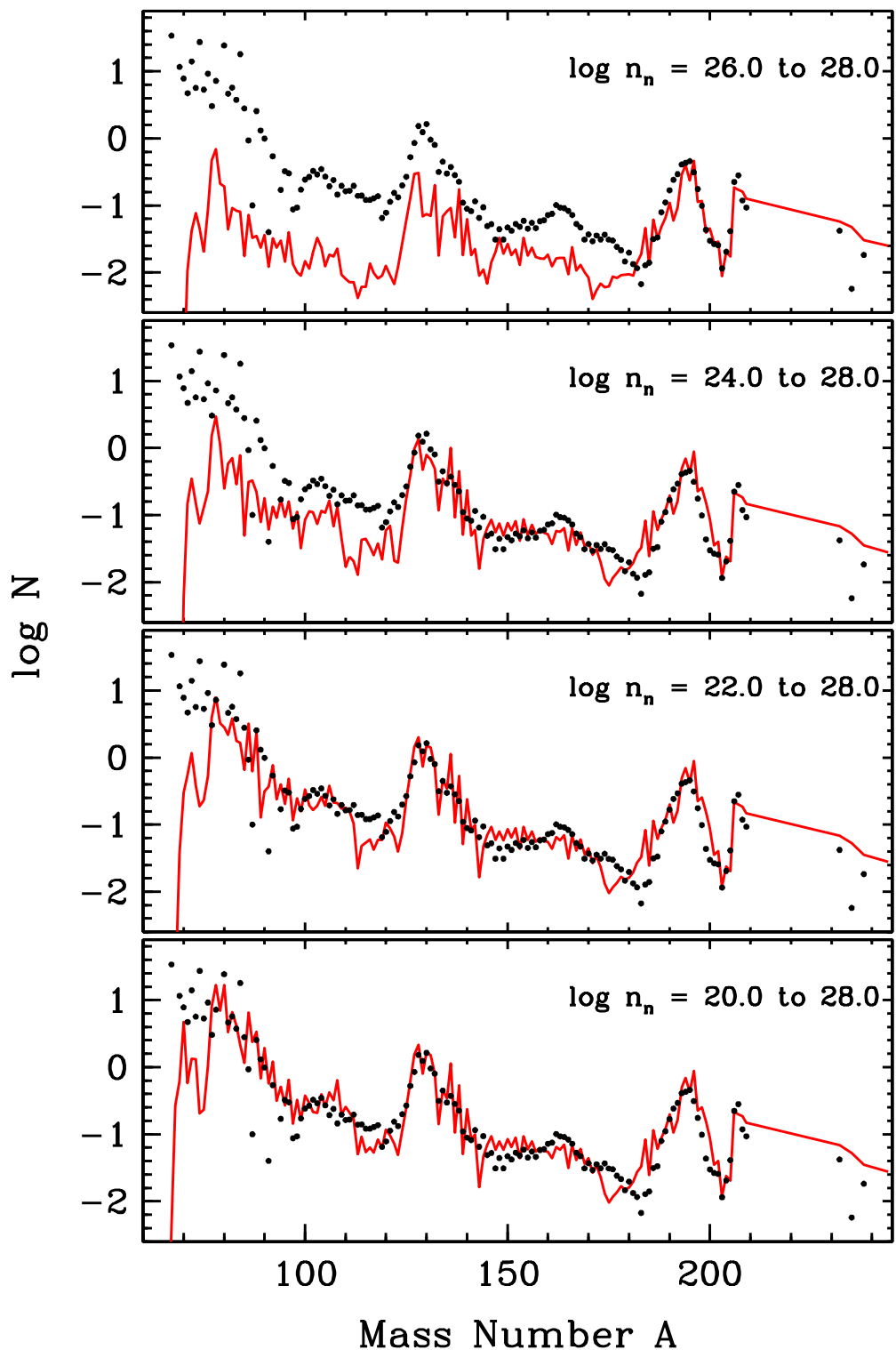}
\caption{
Another comparison of meteoritic \rpro\ abundances and weighted
sums of \rpro\ computations.
For this figure additions commence at the high-density end, and 
progressively add components of lesser values of log~$n_n$ to
fit the lighter \rpro\ abundances.
The symbols and lines are as in Figure~\ref{f5}
\label{f6}}
\end{figure}

Summing the individual isotopic abundances into elemental abundances 
also reproduces the observed SS \rpro\ {\it elemental} 
curve from Z~=~30--82. 
Since we and many other investigators (cited in \S1) have argued that
the abundance pattern in \rpro-rich metal-poor stars is also consistent 
with this SS distribution (Figure~\ref{f1}), it is not 
surprising that multiple neutron density ranges are required to adequately 
match the entire \ncap\ abundance ranges of the stars as well.

\begin{figure}
\epsscale{0.80}
\plotone{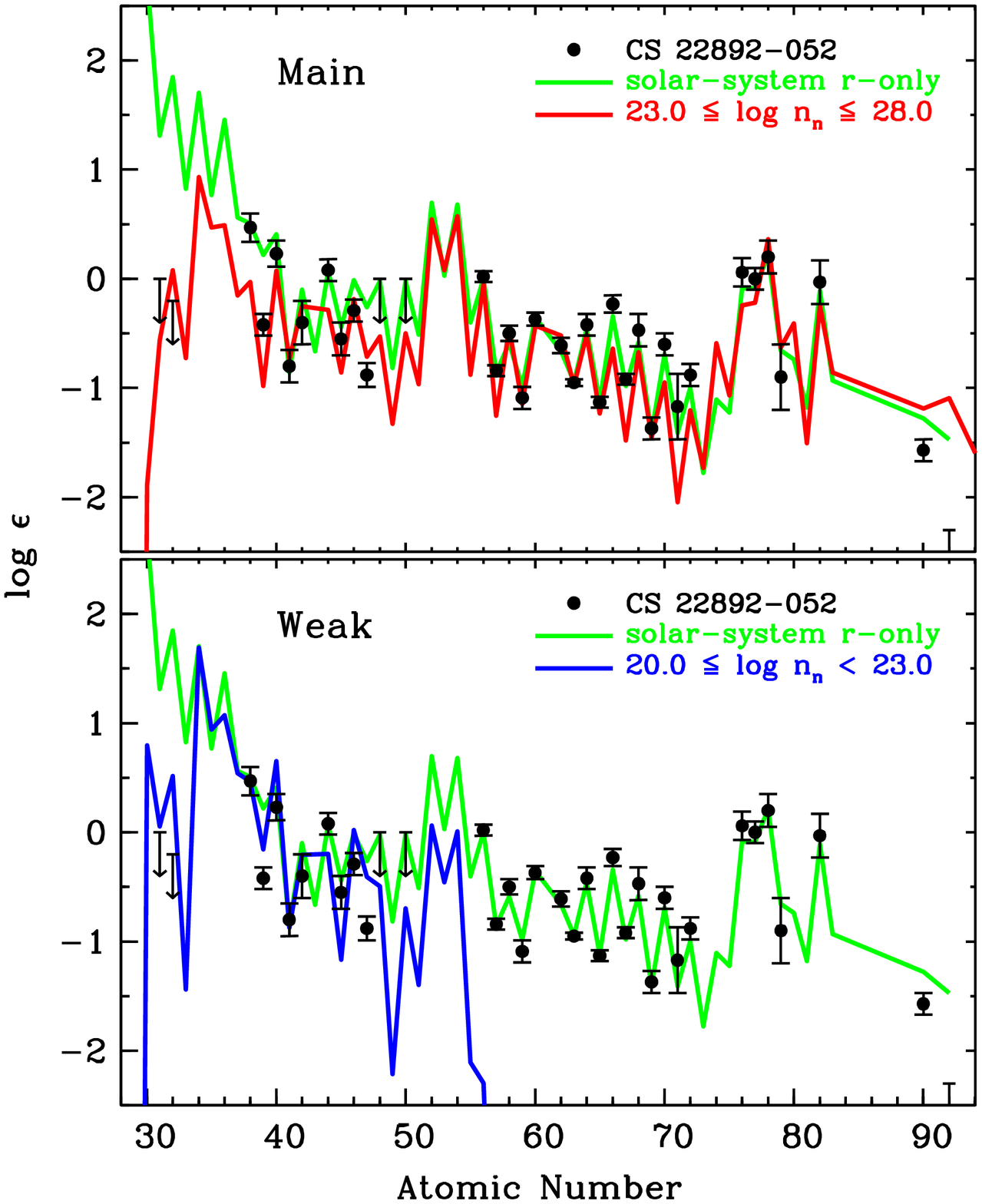}
\caption{
Abundances of \ncap\ elements in \cs22\ compared with \rpro\ distributions.
In both panels the filled circles are the \cs22\ data, mostly from Sneden 
\etal\ (2003), with some updated abundances based on recent transition 
probability studies:
Nd (Den Hartog \etal\ 2003); Sm (Lawler \etal\ 2006a); 
Gd (Den Hartog \etal\ 2006); Hf (Lawler \etal\ 2006b); and
Pt (Den Hartog \etal\ 2005).
Upper limits are shown as downward-pointing arrows.
Also in both panels the SS \rpro\ elemental abundances 
(Simmerer \etal\ 2004), scaled to match the \cs22\ Eu value, are indicated 
by green lines.
In the top panel the ``main'' \rpro\ predictions (neutron number densities
log~$n_n$~$\geq$~23) summed by element are shown with a red line and
normalized  to the stellar Eu abundance.
In the bottom panel, predictions for the ``weak'' \rpro\ are shown with
a blue line where the 
scaling is approximate for illustration purposes.
\label{f7}}
\end{figure}

As an example, we compare in Figure~\ref{f7} the empirical SS 
\rpro\ elemental distribution, theoretical \rpro\ predictions, and 
abundances of the \rpro-rich low metallicity star \cs22\ (Sneden \etal\ 
2003, with updated abundances of Nd, Sm, Gd, Hf, and Pt as indicated 
in the figure caption.)  
The SS distribution has been scaled to match the \cs22\ Eu abundance.
In the top panel of Figure~\ref{f7} the ``main'' \rpro\ abundance 
calculations, for log~$n_n$~$\geq$~23.0, have also been normalized at
the stellar Eu abundance.
Good agreement is seen between the calculations and the mean observed 
abundance levels for stable \ncap\ elements throughout the range
56~$\leq$~Z~$\leq$~82.
Comparing the \cs22\ and the scaled solar abundances (for Z~=~56-82) 
we find an average difference of $<$(\cs22)--SS$_{r-only}$$>$~= 0.044 
and $\sigma$~= 0.096.
The comparison between the stellar and our calculated  \rpro\ abundances 
yields $<$(\cs22)--(main \rpro\ theory)$>$~= 0.15 and $\sigma$ = 0.25.
The main \rpro\ as defined here reproduces the \cs22\ data of the heavier 
elements (Z~$>$~56), including the full A~=~130 \second\ \rpro\ peak, 
including iodine. 
However, both the SS \rpro\ only curve and our calculations (for the 
``main'' \rpro) do not fit the lighter $n$-capture element data 
(Z~$<$~56) in this star.
In particular Y, Mo, Pd and Ag all deviate from both of these curves. 
These differences in fact have been one of the main supports for the 
existence of two $r$-processes. 
We note, however, that the main \rpro\ calculations reproduce the 
odd-Z, even-Z abundance staggering in both light and heavy \ncap\ 
elements in \cs22. 
Further, the calculated abundances fall off significantly with respect 
to the SS $r$-only abundances at the lower atomic numbers, and do not 
seriously clash with the four abundances for which only upper limits have 
been determined: Ge, Ga, Cd, and Sn.

Elemental abundance predictions of the ``weak'' \rpro\ (log~$n_n$~$<$~23)
are overlayed with the observed abundances of \cs22\ 
in bottom panel of Figure~\ref{f7}. 
These predictions only represent the light ``missing part'' of the solar 
\rpro\ abundances (similar to that of the top panel of Figure~\ref{f5}).
The scaling to the \cs22\ light \ncap\ abundances is approximate, for
display purposes only.
Consistency between predicted and observed abundances of these elements
can be achieved under the weak \rpro\ conditions with little
production of the heavier elements (5--10\% of the \second\ \rpro\ 
peak and essentially 0\% for Ba and beyond).  
As can be seen in the figure, the weak \rpro\  predictions do not 
contribute to the heavier (Z$>$ 56) $n$-capture element abundances.
In other words the similar  \rpro\ abundance pattern (for the elements 
with Z $\ge$ 56) seen in the \rpro-rich, metal-poor stars (Figure~\ref{f1}) 
does not need or have any contribution from the weak \rpro. 
The figure also confirms our results from above - that the \first\ SS 
\rpro\ abundance peak can be reproduced with a low $n_n$ ($<$ 10$^{23}$).

The general conclusions about a separation in abundance levels (and 
the associated synthesis conditions) between the lighter and heavier 
\ncap\ elements are  not new, and have been discussed previously 
in the literature (\eg, Sneden \etal\ 2003\nocite{sne03}; 
Ivans \etal\ 2006\nocite{iva06}; Cowan \& Sneden 2006\nocite{cow06}).
However, our new calculations suggest that the separation occurs below 
the element iodine, which appears to be formed along with Ba in the 
A~=~130 abundance peak. 
We note in this regard that there are experimental nuclear data in the 
the \rpro\ path in the mass range from \iso{Fe}{68} to \iso{Te}{140}. 
For these isotopes where the \spro\ data might not be well defined 
(leading to large uncertainties in the \rpro\ residuals), the 
experimental \rpro\ data might be used in the future to directly predict, 
or at least constrain, the actual SS \rpro\ abundances.
Such data could be employed to quantify where in mass (atomic number) the
possible separation between the main and weak \rpro\ (if one exists) occurs.


\section{DETAILED ABUNDANCE SIGNATURES OF THE $r$-PROCESS}

In this section we consider some additional \rpro\ abundance clues
from the details of our calculations, beyond considerations of the overall
fits to solar and stellar abundance distributions. In the near future
some isotopic, as well as additional (yet unobserved) 
elemental, abundance ratios will become available
for certain stars. Such ratios will provide increasingly stringent
constraints on both $s$- and \rpro\ nucleosynthesis contributions. 
Our calculations predict isotopic abundance ratios for the light-heavy 
elements -- up to the heavy wing of the A=130 peak which includes for 
example the Ba isotopes. 
At least within the high entropy model, the freezeout is very fast; 
non-equilibrium-captures do not play a significant role; $P_n$ values
are small (Kratz \etal\ 1993\nocite{kra93}; Freiburghaus \etal\ 
1999\nocite{fre99b}; Rauscher 2004\nocite{rau04}; Farouqi \etal\ 
2006\nocite{far06}). 
Hence, the final abundance pattern of the {\it whole} A~=~130 peak, 
including the Ba isotopes, is mainly determined by the initial 
``progenitor abundances'' with their odd-Z/even-Z staggering smoothed 
out by delayed neutron emission during the decay back to stability.  
These calculations should not be viewed as necessarily tightly 
constraining those isotopic ratios, but instead predicting a reliable
range of values as a function of neutron number densities or
entropy.


\subsection{Barium Isotopic Fractions}


An additional probe of the heavy element pattern and its range in metal-poor 
stars is provided by the (limited) isotopic abundance information that has 
been reported in the literature.
The isotopes of Eu in metal-poor halo stars have been investigated
by Sneden \etal\ (2002)\nocite{sne02} and Aoki \etal\ 
(2003a,b).\nocite{aok03a,aok03b} 
For several \rpro-rich stars, the isotopic abundance fraction ratios, 
$f(\iso{Eu}{151})/f(\iso{Eu}{153})$~= N(\iso{Eu}{151})/N(\iso{Eu}{153})~=
1.0~$\pm$~0.1, are in good agreement with the meteoritic \rpro\ fraction,
$f(\iso{Eu}{151})/f(\iso{Eu}{153})$~= 0.478/0.522~= 0.916 (Anders \& 
Grevesse 1989).\nocite{and89}
Estimates of the photospheric Eu isotopic ratio (Hauge 1972\nocite{hau72}, 
Lawler \etal\ 2001\nocite{law01}) are also in accord with these values.

Barium has five naturally-occurring isotopes, A~=~134--138, which are 
produced in substantially different amounts in the $s$- and \rpro.\footnote{
Two other very minor but stable isotopes, \iso{Ba}{130} and \iso{Ba}{132}, 
are products of the $p$-process and not relevant to the present study.}
Even-Z isotopes \iso{Ba}{134} and \iso{Ba}{136} cannot be reached 
by the \rpro.  
As a result the abundance ratio of odd-A isotopes to the total, 
defined as $f_{odd}$~$\equiv$ [N(\iso{Ba}{135})+N(\iso{Ba}{137})]/N(Ba),
is larger in \rpro\ than \spro\ nucleosynthesis events; 
$f^r_{odd}$~$>$~$f^s_{odd}$.
In principle observation should be able to assess $f_{odd}$ in stars, 
for the odd-Z isotopes \iso{Ba}{135} and \iso{Ba}{137} have hyperfine-split 
line substructures that the even-Z isotopes lack.
Therefore, careful measurement of the line broadening of \ion{Ba}{2} lines
should indicate the relative $r$-/$s$-process contributions to Ba production.

Unfortunately even the SS value of $f^r_{odd}$ for Ba, which can 
be determined from meteoritic studies, is not known to high accuracy.
This stems from the overall \spro\ dominance of barium in SS material.
The \rpro\ contributions to the barium isotopes are small, thus any 
isotopic abundance uncertainties are magnified in the estimates of their
\rpro\ contributions.
Lodders (2003)\nocite{lod03} has given a new empirical estimate of the 
\rpro\ and \spro\ contributions to each barium isotope, and with those 
data we compute a value of  $f^r_{odd}$~=~0.72.
The Lodders isotopic breakdowns are very similar to those that were
used by Burris \etal\ (2000)\nocite{bur00} and Simmerer \etal\ 
(2004)\nocite{sim04} to compute total solar elemental abundances of 
\ncap\ elements (see also the isotopic breakdowns in Cowan \etal\ 
2006).\nocite{cow06b}
Arlandini \etal\ (1999)\nocite{arl99} subtracted \spro\ model calculations 
from the solar isotopic abundances to yield their \rpro\ contributions. 
From their data we compute $f^r_{odd}$~=~0.46.  

These independent assessments of $f^r_{odd}$ for Ba in the solar system
are in agreement mainly because the uncertainties in each are large, 
about $\pm$0.2.
Lacking a clear indication of which value to adopt, we simply average them
to adopt $f^r_{odd}$~$\approx$ 0.60~$\pm$~0.20.
In contrast $f^s_{odd}$ is tightly constrained between values of 0.09 to 0.11 
in all of the estimates, but this again is due to the \spro\ dominance of 
Ba synthesis in SS material.
From these independent assessments we adopt $f^s_{odd}$~= 0.10~$\pm$~0.02.

\begin{figure}
\epsscale{0.9}
\plotone{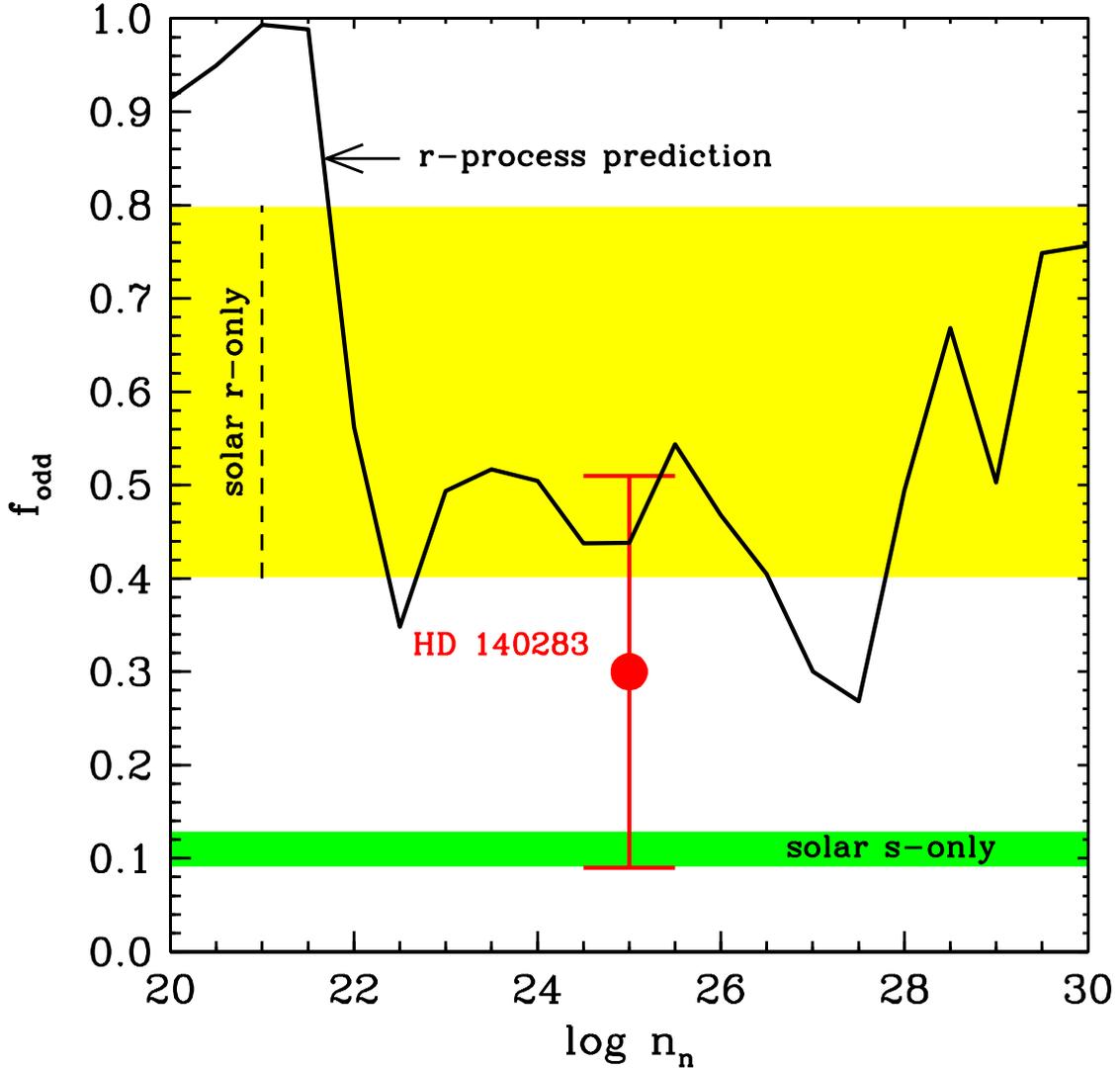}
\caption{
Predicted \rpro\ barium odd-Z abundance fractions,
$f^r_{odd}~=~[N(\iso{Ba}{135})~+~N(\iso{Ba}{137})]/N(Ba)$, 
plotted versus neutron number density.
In addition to the black solid curve representing the predicted values,
we have drawn a red solid circle with error bar to indicate the 
observed $f_{odd}$~= 0.30~$\pm$~0.21 in the very metal-poor halo star 
HD~140283 (Lambert \& Allende Prieto 2002).
This point has been arbitrarily placed at log~$n_n$~=25 for illustration
in the figure, but only the vertical placement of the point has meaning
here.
Also shown is a yellow color band indicating the SS \rpro-only 
$f^r_{odd}$~= 0.6~$\pm$~0.2 range, and a green band indicating the 
\spro-only value $f^s_{odd}$~= 0.11~$\pm$~0.02 range.
\label{f8}}
\end{figure}

In the Table~\ref{tab1} we list our theoretical \rpro\ predictions 
for Ba isotopic abundances, and the resulting $f^r_{odd}$ values.
These calculations are for single fixed neutron number density conditions 
for 20~$<$~log~$n_n$~$<$~30.
We show the trend of $f^r_{odd}$ with neutron number density in 
Figure~\ref{f8}, indicating also in the figure the SS
values of $f^r_{odd}$ and $f^s_{odd}$.
In addition to the theoretical calculations adopted here, some  
full dynamic network calculations have been performed (Farouqi 
\etal\ 2006)\nocite{far06} in the context of the high-entropy 
supernova model -- with post processing during the non-equilibrium 
neutron-freezeout phase.
These computations have been made to check or constrain
the more simplistic, but detailed, parametric predictions from the 
waiting-point approximation employed here.
The Farouqi \etal\ calculations employ an extension of the dynamic \rpro\ 
code (Freiburghaus \etal\ 1999b\nocite{fre99b}; 
Cowan, Thielemann \& Truran 1991\nocite{cow91}), but
include \ncap\ cross sections from the NON-SMOKER code 
(Rauscher \& Thielemann 2000\nocite{rau00}).
For a range of conditions (electron abundance $Y_e$ from 0.41--0.49, 
entropy $S$ of 200-300 k$_b$/baryon) these dynamic calculations
predict a range of $f^r_{odd}$ values of 0.48 to 0.51, thus
overlapping with the waiting-point approximation values.

Few assessments of $f_{odd}$ in metal-poor stars have been attempted.
The wavelength shifts due to isotopic mass differences and 
hyperfine splits for the two odd-A barium isotopes are comparable to the 
line widths (due to thermal and microturbulent broadening) in stellar 
spectra, making isotopic abundance determinations very challenging.
Recently, Lambert \& Allende Prieto (2002)\nocite{lam02}, following 
an earlier investigation by Magain (1995), observed the \ion{Ba}{2} 
4554~\AA\ line in the halo subgiant HD~140283 ([Fe/H]~= --2.4), at very 
high spectral resolution and signal-to-noise.
Detectable excess breadth of this line compared with single-component
absorption features would be due mainly to the hyperfine splitting of the 
odd-A isotopes. 
Lambert \& Allende Prieto detected the excess broadening at a marginal level,
and their line profile analysis yielded a combined fractional abundance of 
the odd-A isotopes $f_{odd}$ =  0.30~$\pm$~0.21.

The HD~140283 result is shown in Figure~\ref{f8}.
Within large uncertainties, $f_{odd}$(HD~140283)~$\sim$ $f^r_{odd}$(s.s.).
Frustratingly, the range in isotopic values for HD~140283
does not permit exclusion of the SS \spro\ value.
Given that the HD~140283 result is marginally more consistent with an 
\rpro\ than an \spro\ origin for that star, we examine the implications of 
our \rpro\ calculations.
As indicated in Table~\ref{tab1} and Figure~\ref{f8},
the lowest neutron number densities, log~$n_n$~$\lesssim$~22, produce
very small amounts of Ba.
Since these low densities also yield $f^r_{odd}$ values at odds with
SS values, they are excluded from consideration here.
Likewise, the highest neutron number densities, log~$n_n$~$>$~28, predict
Ba isotopic fractions that are much larger than the values permitted
by the HD~140283 measurement.
Averaging the isotopic abundances in the range 
23~$\leq$~log~$n_n$~$\leq$~28 yields $f^r_{odd}$~=~0.44, very similar 
to the Arlandini \etal\ (1999) value and that found from the preliminary 
dynamic calculations. 
The meager Ba isotopic abundance data do not permit a more sharply defined
neutron density range.


\subsection{Barium and Europium $r$-Process-Rich Metal-Deficient Stars}


More progress can be made by considering Ba/Eu elemental abundance
ratios, which have been used for decades to estimate the relative influence
of \spro\ and \rpro\ contributions to stellar \ncap\ abundances
(\eg, see Spite \& Spite 1978).
For SS material, log~$\epsilon_{r}$(Ba)~= +1.446 and
log~$\epsilon_{r}$(Eu)~= +0.494, or
log~$\epsilon_{r,s.s}$(Ba/Eu)~= +0.952.
For HD~140283 Gratton \& Sneden (1994) reported [Ba/Fe]~= $-$0.64~$\pm$~0.06
and [Eu/Fe]~= +0.09~$\pm$~0.01, yielding [Ba/Fe]~= $-$0.73~$\pm$~0.06,
which translates to $<$log~$\epsilon_{obs}$(Ba/Eu)$>$~= +0.92~$\pm$~0.06 
using their solar Ba and Eu abundances.
This abundance ratio is obviously in good agreement with 
log~$\epsilon_{r,SS}$(Ba/Eu), strengthening the suggested attribution 
of $f_{odd}$(Ba) to an \rpro-only synthesis.

\begin{figure}
\epsscale{0.9}
\plotone{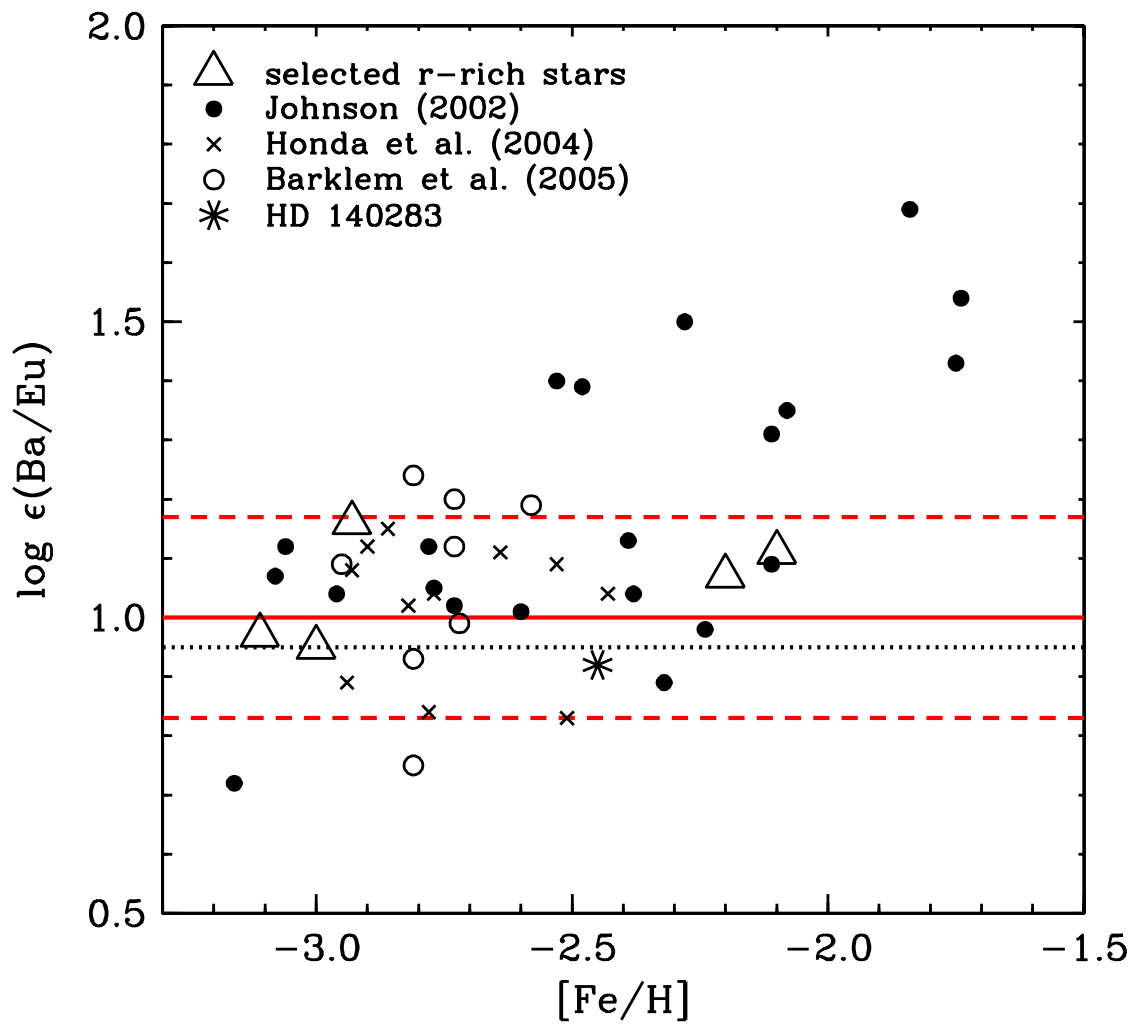}
\caption{
Observed barium-to-europium abundance ratios in metal-poor stars.
The data shown here are representative of the recent literature,
and their sources are given in the text.
The black dotted line represents the solar \rpro\ value of 
log~$\epsilon$(Ba/Eu).
The red solid line with the two red dashed lines indicate our best estimate 
of this ratio and its uncertainty in \rpro-rich stars.
\label{f9}}
\end{figure}

Many very low metallicity stars exhibit similar Ba/Eu ratios.
We surveyed the  literature to estimate these spectroscopic values.
In Figure~\ref{f9} we plot [Ba/Eu] as a function of [Fe/H] from 
several recent halo-star comprehensive abundance analyses: 
Johnson (2002)\nocite{joh02}, Honda \etal\ (2004)\nocite{hon04}, and 
Barklem \etal\ (2005).\nocite{bar05}
We added points from detailed studies of a few extremely
\ncap-rich stars (Westin \etal 2000\nocite{wes00}, Cowan \etal\ 
2002\nocite{cow02}, Hill \etal\ 2002\nocite{hil02}, Sneden \etal\ 
2003\nocite{sne03}, and Ivans \etal\ 2006\nocite{iva06}).
This is a representative (but not complete) list of Ba/Eu abundance
studies in this metallicity domain.
A large majority of the stars considered in the cited studies exhibit
$<$log~$\epsilon_{obs}$(Ba/Eu)$>$~$\simeq$ +1.00.
Some points scatter to significantly larger values ($\gtrsim$+1.4)
of this quantity, clearly indicative of substantial \spro\ contributions 
to Ba in those stars.
The envelope containing essentially the entire star-to-star scatter of the 
Ba/Eu ratios observed in \rpro-rich stars is $\approx$0.35 in width, or
$<$log~$\epsilon_{obs}$(Ba/Eu)~$>$~= +1.00~$\pm$~0.17

\begin{figure}
\epsscale{0.9}
\plotone{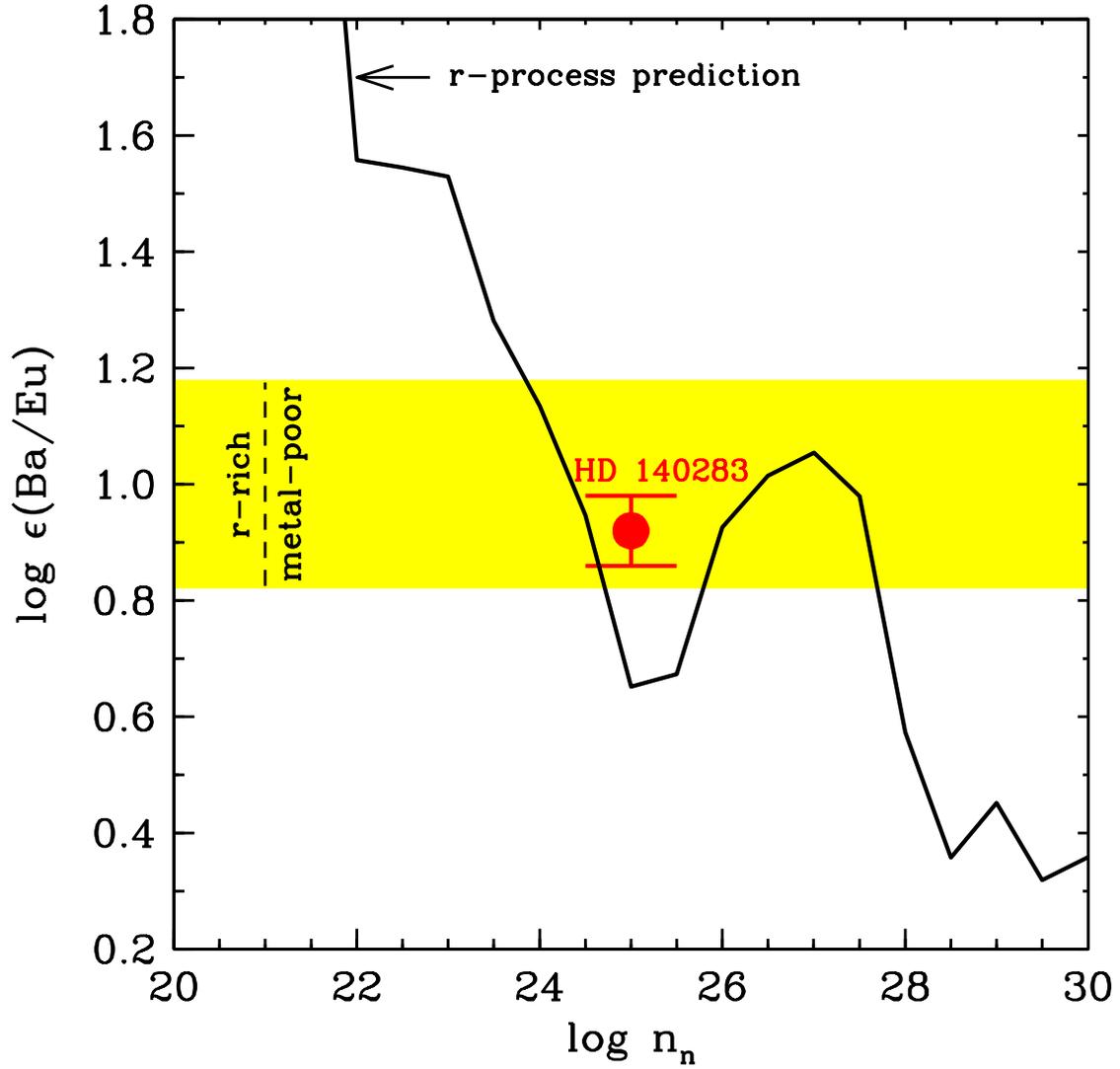}
\caption{
Predicted elemental abundance ratios log~$\epsilon$(Ba/Eu) plotted 
versus neutron number density log~$n_n$.
The red point with error bar indicates the value derived for HD~140283 by 
Gratton \& Sneden (1994). 
Also shown is a yellow band covering the log~$\epsilon$(Ba/Eu) range 
reported in the literature for \rpro-rich very metal-poor stars, as
established in the previous figure.
\label{f10}}
\end{figure}

In Figure~\ref{f10} we show our calculated Ba/Eu abundance 
ratios plotted as a function of neutron number density.
To this figure we add a point representing the value for HD~140283, and
a color band representing the range in values for \rpro-rich 
low metallicity stars.  
The neutron number density range containing theoretical predictions and
stellar observations is substantially less than that inferred from 
consideration of Ba isotopic abundances alone.
This figure, together with log~$\epsilon_{r,s.s}$(Ba/Eu)~= +0.95, 
suggests that 24~$\lesssim$ log~$n_n$~$<$ 28, with fairly well-defined 
boundaries on both ends of the allowable neutron number density range.
Densities beyond this domain appear unable to reproduce $f_{odd}$(Ba)
and log $\epsilon$(Ba/Eu) of metal-poor \rpro-rich stars, and by extension
the relative abundances of most rare-earth elements in these stars.


\subsection{The Iodine--Barium Connection}


The existence of two distinct \rpro\ synthesis sites began with 
the Wasserburg \etal\ (1996) critical assessment of \iso{I}{129} 
and \iso{Hf}{182} meteoritic abundance levels.
In the preceding subsections we demonstrated that \rpro\ computations 
with an addition of neutron densities in the range 
24~$\lesssim$ log~$n_n$~$<$ 28 yields both barium isotopic
and Ba/Eu elemental abundance ratios that are compatible with the 
solar \rpro\ abundances.

\begin{figure}
\epsscale{0.9}
\plotone{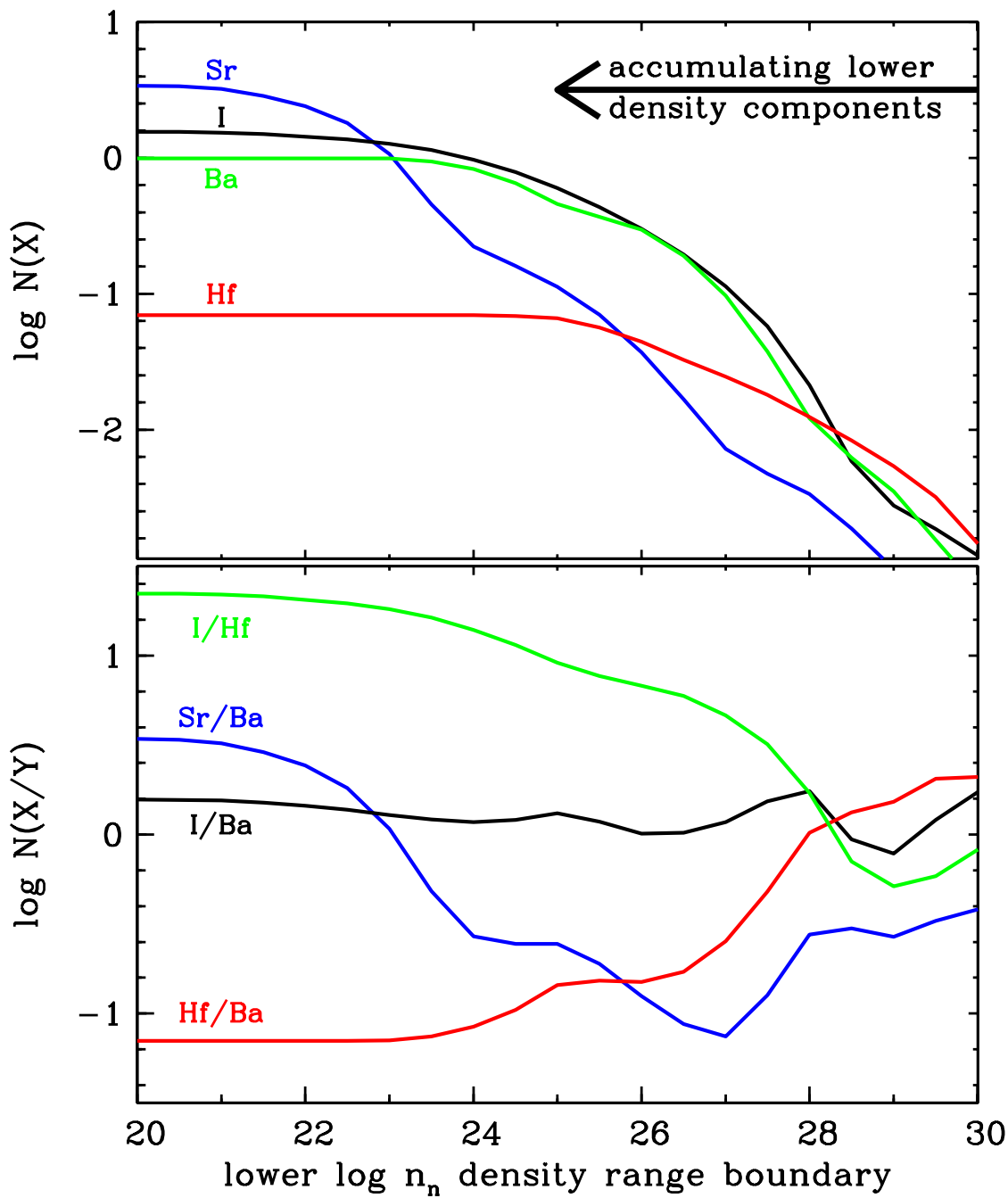}
\caption{
Abundance variations of Sr, I, Ba, and Hf as a function of neutron number
density range.
N is the abundance on
the standard meteoritic scale where log N$_{\rm Si}$~=~6.
In the top panel the weighted abundance accumulations of each element 
are plotted.
These are computed starting with the predicted abundances at the
highest neutron density, log~$n_n$~=~30, and adding the abundances
at successively lower values of log~$n_n$. 
The abundances of particular elements ``saturate'' at their SS 
\rpro\ numbers, when the contributions of even smaller neutron density 
regimes cease to add to their abundances.
In the bottom panel, the variations of selected ratios of the abundances
are displayed.
\label{f11}}
\end{figure}

Computed abundances of Sr (Z~=~38), I (53), Hf (72), and Ba (56), 
are shown in the top panel of Figure~\ref{f11} and several abundance
ratios among these elements are shown in the bottom panel.
These are plotted as a function of the neutron density range defined
in the following way.
Beginning at the highest neutron density log~$n_n$~=~30, successively 
smaller neutron density steps represent the accumulation of the weighted 
abundances resulting from that particular neutron density {\it plus} the 
total of all higher densities.
The highest values of log~$n_n$ contribute only trace amounts of the
solar and stellar \rpro\ abundances, which then grow as smaller neutron
density contributions are added into the sums.
The abundance curve of the heavy element Hf ``saturates'' at its 
approximate solar value most quickly (at log~$n_n$~$\simeq$ 25.5) in 
agreement with the overall trend shown in Figure~\ref{f6}, while the light 
element Sr does not reach its SS $r$-process value until log~$n_n$~$\simeq$ 21.5.

What is more important in Figure~\ref{f11} is the close tracking 
between I and Ba in our computations.
These neighboring elements correlate over a wide neutron density range -- 
the slope of their abundance ratio shown in the bottom panel of 
Figure~\ref{f11} is essentially flat. 
The situation for I and ``distant'' Hf ($\Delta$Z~=~19) is somewhat different. 
Here, large differences (factors of two or more) occur for ``high'' 
neutron densities, log~$n_n$~$\gtrsim$ 25.
These also are conditions where the Ba abundance is no longer solar.
For conditions of a ``main'' \rpro\ with full solar I, Ba and 
Hf $r$-abundances (\ie, 23~$<$~log~$n_n$~$<$~28),
iodine at the top of the A~=~130 N${_r,\odot}$ peak and hafnium at 
the onset of the A~=~195 N${_r,\odot}$ peak are ``coupled''
([I/Hf]$_{r,\odot}$ $\simeq$ 0.85). 
Thus, for the conditions in the main \rpro\ I, Ba  and Hf appear to be 
synthesized together, and in solar proportions.  
Sizable abundance ``decouplings'' of these three elements only occur at 
the highest neutron densities, log~$n_n$~$\gtrsim$~25, when 
[I/Hf] and [Hf/Ba] do not retain their solar \rpro\ abundance ratios.

For ``main'' \rpro\ conditions, 23~$<$~log~$n_n$~$<$~28, we estimate 
that both the abundance level of iodine and the (I/Hf) ratio are 
approximately 90\% of their SS \rpro\ values. 
In the context of our \rpro\ model, this conclusion seems quite robust 
and results from two primary factors.  
First, from nuclear-structure arguments there is a bottle-neck
behavior in the \rpro\ flow at the N~=~82 shell-closure.
Thus,  full-solar $r$-process Ba production  is accompanied by the full-solar 
A~=~130 abundance peak synthesis -- at least at its top.
Second,  the ``classical'' \rpro\ reaches (or enters) the N~=~82 shell at 
{\it lower} Z than the top of the peak (\iso{Ag}{129} or \iso{Cd}{130}). 
Our parameterized \rpro\ waiting-point predictions, as well as the 
preliminary Farouqi \etal\ (2006)\nocite{far06} dynamic calculations, 
show that N~=~82 is already reached in the Tc (Z~=~43) isotopic chain 
-- in the very extreme even in the Zr (Z~=~40) chain. 
Hence, the ``dividing line'' between the two $r$-processes appears to fall
well below the \iso{Ag}{129},  because of the N~=~82 shell closure far 
from stability.  
Instead,  \iso{Ag}{129} is included with the ``rising wing'' of the 
A~=~130 \rpro-abundance peak from about A~=~125 upwards.


\subsection{Hafnium and the Third $R$-Process Peak}


Hafnium (Z~=~72) is the next element beyond the rare-earth group, and it may 
serve as an important link between those elements and the \third\ \rpro-peak.
For practical abundance determinations, we note that Hf, like
all of the rare earths and the radioactive elements Th and U, is
detectable in metal-poor stars via absorption lines arising from low
excitation states of the first ion (see Lawler \etal\ 2006b).\nocite{law06b}
Therefore, observed ratios of Hf/Ba, Hf/Eu, Hf/Th, etc., are very
insensitive to uncertainties in stellar atmospheric parameters 
T$_{\rm eff}$ and log~$g$.  
This is a happier situation than that existing for ratios 
involving \third\ \rpro-peak elements, such as Eu/(Os,Ir,Pt) or 
Th/(Os,Ir,Pt), because the \third-peak elements are detectable only via 
their neutral species.
Atmospheric parameter errors are echoed directly into such ratios.
Additionally, the dominant transitions of \third-peak neutral species 
lie in the UV ($\lambda$~$<$~4000~\AA), making detections and reliable
abundance determinations for these elements difficult. 
 
\begin{figure}
\epsscale{0.9}
\plotone{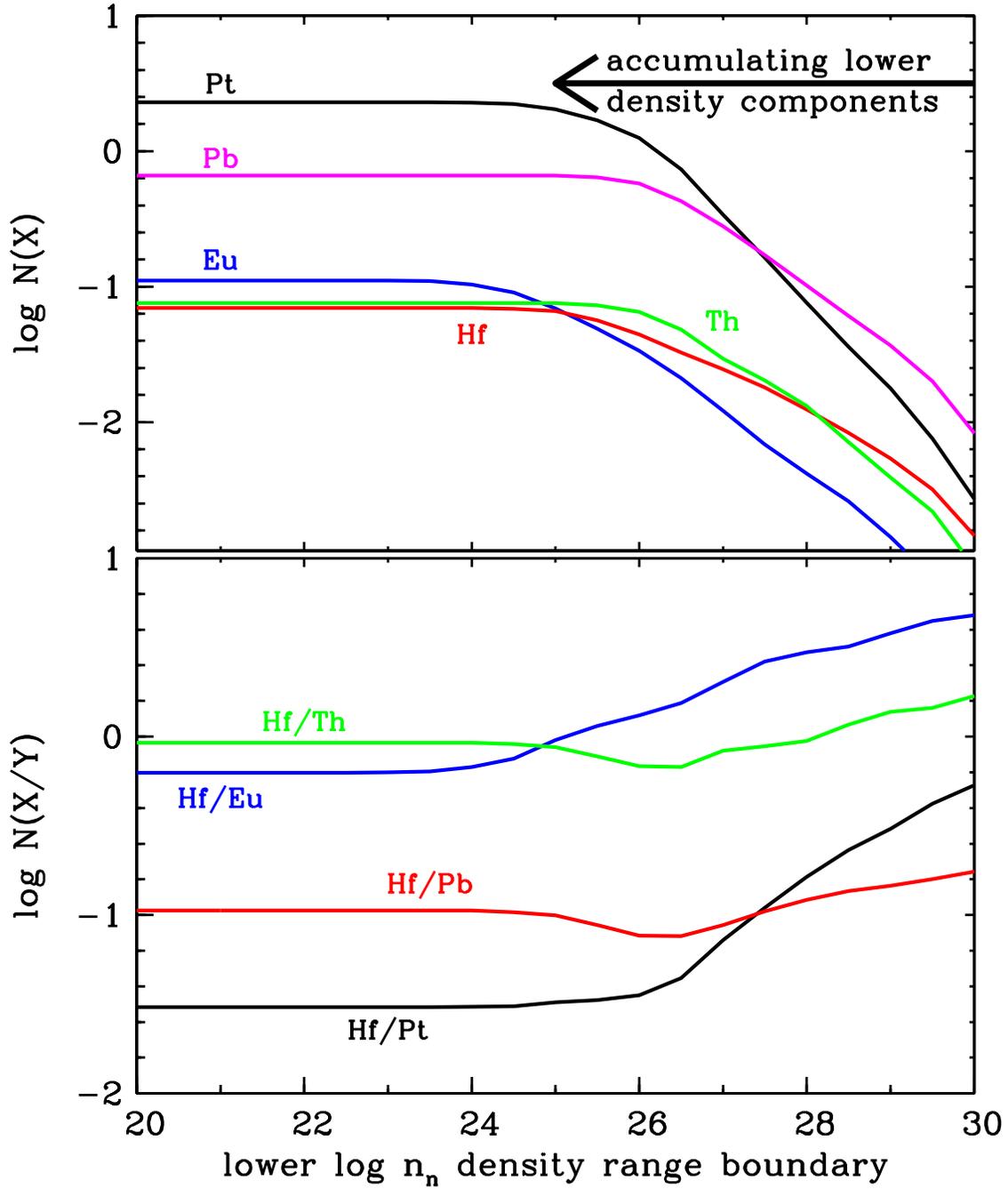}
\caption{
Abundance variations of Eu, Hf, Pt, Pb, and Th as a functions of neutron
number density range.
The meanings of the curves are as in Figure~\ref{f11}.
\label{f12}}
\end{figure}

In Figure~\ref{f12} we show abundance variations of Eu, Hf, Pt, Pb,
and Th as a function of neutron density range, in the same fashion as
was done for lighter elements in Figure~\ref{f11}.
Variations in abundance ratios of these elements (shown in the bottom panel 
of Figure~\ref{f12}) are small for neutron densities log~$n_n$~$<$~28.
Note in particular the near constancy of the Hf/Th ratio over the entire
neutron density range of the \rpro.
Our calculations suggest that spectroscopists should invest effort
in the determination of Hf abundances (using such \ion{Hf}{2} transitions
as those at 4093.15 and 3918.09~\AA) to see if the ratios of Hf/Eu and
Hf/Th are constant in \rpro-rich metal-poor stars.
If so, this may strengthen the use of Th/Hf  in cosmochronometry studies.


\section{\bf NUCLEAR CHRONOMETERS}


In this paper we have emphasized the robustness of the \rpro\ production 
of heavy nuclei in the mass range from A~$\simeq$~130 through the actinide 
region, as reflected in the observed abundance patterns for metal-deficient 
\rpro-enriched field halo stars. 
Our ability to identify this pure \rpro\ pattern is a consequence 
of the fact that one can identify low metallicity stars for which no 
significant \spro\ contamination from longer lived asymptotic giant branch 
stars (Busso \etal\ 1999) has yet occurred. 
An important further measure of this robustness is provided by the general 
consistency of the stellar ages obtained with chronometers. 
We will examine specifically both the case of the 
classical thorium/uranium actinide chronometer and that of the 
thorium/europium chronometer.

We consider again the abundance distributions of five ``pure'' \rpro\ 
stars shown in Figure~\ref{f1}, concentrating now on the
the actinide chronometers \iso{Th}{232} (hereafter simply called Th), 
\iso{U}{235}, and \iso{U}{238} (which can be formed {\it only} in the \rpro).
We have computed \rpro\ abundances using various global mass models,
constraining all of them to reproduce the SS stable \rpro\
pattern, with special emphasis on fitting the \rpro\ yields of the 
\third\ peak and extrapolating to masses A~$\simeq$~250.
In Table~\ref{tab2} we summarize these computations.
Column~2 lists abundance ratios derived from calculations described
in Cowan \etal\ (1999)\nocite{cow99}.
Columns~3 and 4 are these ratios from new \rpro\ calculations using
Fe seeds that yield the best overall fit to the stable abundance data 
for masses A~$>$~83 (``fit1'', considering all \ncap\ isotopes), 
and for masses A~$>$~125 (``fit2'', restricted to only those isotopes
matched by the main \rpro).
Column~5 employs similar calculations using a Zr seed that best match
the A~$>$~125 nuclei.
Finally, column~6 lists the Th abundance from the fit2 calculation to
the present-day observed SS elemental abundances.

Table~\ref{tab2}'s Th/U production ratios lie in the range
1.475~$\leq$~Th/\iso{U}{238}~$\leq$~1.805.
The present computations suggest Th/\iso{U}{238}~$\simeq$~1.5 
for this important chronometer pair.
The new lower values result from improved nuclear data.
Specifically, there is a significant change in the \rpro-matter flow 
through the A~$\simeq$~130 bottle-neck region of the N$_{r,\odot}$
peak, which continues to affect the build-up of the third r-peak and the
formation of the heaviest r-elements. 
Therefore slightly less material is shifted beyond Bi than in our earlier 
approach (Pfeiffer \etal\ 1997\nocite{pfe97}, 
Cowan \etal\ 1999\nocite{cow99}), yielding somewhat lower
Th/\iso{U}{238}, \iso{U}{238}/\iso{U}{238}, and Th/\third-peak 
abundance ratios. 

We have taken into account in a crude manner the main fission modes,
spontaneous and neutron-induced fission, in the calculations. 
We make the simplifying assumption that everything beyond mass number 
256 (\ie, \iso{Cf}{256}) fissions completely, analogously to what
we have done previously (\eg, Cowan \etal\ 1999).\nocite{cow99}
Furthermore, we have verified again that the known nuclei undergoing 
spontaneous fission for A~$<$~256 make no significant contribution to 
the mass range 232~$<$~A~$<$~255.
For cases below the A~=~256 region, new multi-dimensional fission barrier 
calculations (M\"oller \etal\ 2001\nocite{mol01}, and private communication 
2006) show that spontaneous and neutron-induced fission have no effect,
since the barrier heights of isotopes in the \rpro\ path (with 
$S_n$~$<$2~MeV) are $>$8~MeV.
Also in the $\beta$-decay back to the valley of the mass parabola, the 
isobars still have fission barriers of 6-7~MeV.
Hence, at least from the above model predictions, spontaneous and 
neutron-induced fission should be negligible.
Our (site independent) waiting point approximation calculations, as well 
as dynamical network calculations (Farouqi \etal\ 2006\nocite{far06}
for the SN II high-entropy wind scenario) indicate that the amount of 
matter actually involved in fission in the \rpro\ beyond mass number 
256 is of the order of 1-5\% of the total amount of matter.
The only minor effect in this mass region instead comes from
$\beta$-delayed fission.

However, most all of these calculations up to now neglect possible effects 
from $\beta$-delayed fission ($\beta$df), since they are generally believed 
to be unimportant. 
Nevertheless, we have tried to estimate the $\beta$df-rates of 20 potential 
$\beta$df-candidates between \iso{Fr}{252} and \iso{Am}{277} using
the above QRPA(Quasiparticle Random Phase Approximation) 
model for GT and GT+ff strength functions and the recent 
fission barriers of Mamdouh \etal\ (1998)\nocite{mam98} (based on the 
ETFSI-1 mass model) by following the simple approach outlined by 
Kodama \& Takahashi (1975).\nocite{kod75}
We then have compared these results with the ``complete-damping'' and 
``WKB barrier penetration'' calculations of $\beta$df-rates by 
Meyer \etal\ (1989)\nocite{mey89}, who used the same QRPA
model for GT strength functions and the (systematically lower) fission 
barriers of Howard \& M\"oller (1980).\nocite{how80}
This comparison convinces us that the effect of the $\beta$df-mode on 
the final Th and U abundances is small but may not be completely negligible. 
It may reduce the final Th abundance by about 8\% and the final U abundance 
by approximately 3\%. 
This would reduce our ``fit2'' Th/U abundance ratio from 1.568 
(see Table~\ref{tab2}) to 1.479, which would still lie within the
range of acceptable ratios. 
On the other hand, one might think of including the potential $\beta$df
effects in the uncertainties of the $r$-chronometer ages. 
For example, this would add a further uncertainty of 0.55~Gy 
to the Th/Eu age of a star like \cs22, or to the Th/U age of  
stars like CS 31082-001 or \bd17.


\subsection{\bf The Th/U Chronometer}


We first examine the case for the Th/U chronometer.\footnote{
For this purpose we approximate the uranium abundance as that of $^{238}$U, 
given the significantly shorter half-life of $^{235}$U 
($\tau_{1/2}$ = 7.038x10$^8$ years) compared to $^{238}$U 
($\tau_{1/2}$ = 4.468x10$^9$ years).
The thorium ($^{232}$Th) half-life is $\tau_{1/2}$ = 1.405x10$^{10}$ years.}
This chronometer ratio is considered ideal as the elements are 
\rpro-only and near each other in nuclear mass number.
As we have noted, however, 
the one optically available 
uranium line is very weak in stellar spectra and  
blended with molecular lines, making its detection very difficult.  
Furthermore, for the Th/U ages to be meaningful they need to be determined 
in the context of (or constrained by) the abundances of other stable 
elements, particularly the \third\ \rpro-peak elements and Pb and Bi.  
We consider the conditions we have explored for which we obtain the best 
overall agreement with the observed patterns in halo stars. 
The mass model chosen for this purpose is the ETFSI-Q, 
yielding the ratios shown in Table~\ref{tab2}. 

There are only three halo stars for which there exist observational 
determinations  of both the thorium and the uranium abundances, along
with detailed abundances of many stable elements. 
The observed Th/U ratio for CS 31082-001 (Hill \etal\ 2002\nocite{hil02}) is 
Th/U~=~8.7 (log(Th/U)~=~0.94~$\pm$~0.11), while that for \bd17\ is
Th/U~=~7.6 (log(Th/U)~=~0.88~$\pm$~0.10). 
For these abundance ratios and our production ratio (Th/U)$_0$~=~1.557, 
the ages for the two halo stars CS 31082-001 and \bd17\ are, 
respectively, 16.2~Gyr and 14.9~Gyr, both having uncertainties of 
approximately $\pm$3--3.5 Gyr arising from observational uncertainties. 
Very recently a new uranium detection has been made in
HE 1523-0901 by Frebel et al. (2007), who  
find 
Th/U~=~7.24 (log(Th/U)~=~0.86~$\pm$~0.15) for this star.


\subsection{\bf The Th/Eu Chronometer}


An alternative to the Th/U chronometer is the ratio 
of the abundance of the long lived radioactive Th nucleus to the 
abundance of the stable \rpro\ product europium. 
Eu is formed almost entirely in the \rpro\ (\eg, Simmerer \etal\ 
2004\nocite{sim04} and references therein) 
and is readily observable from the ground.
The Th/U chronometer pair may be more robust and intrinsically more 
accurate, but determinations of both the Th and Eu abundances are  
available for many more halo stars than Th and U. 
It is important, therefore, to quantify the use of this Th/Eu chronometer 
and the reliability of age determinations resulting from it. 

The production ratio (Th/Eu)$_0$ from our current study is 0.530 
(Table~\ref{tab2}). 
This value is slightly higher than previous values used in our age 
calculations (\eg, Cowan \etal\ 1999\nocite{cow99}) and results from 
better nuclear data and better fits to the stellar and solar abundance 
data (see discussion above). 
The observed Th/Eu ratio for \bd17\ is 0.309 (Cowan \etal\ 
2002),\nocite{cow02} leading to an implied age of 10.9~Gyr. 
Previously determined Th/Eu ages for other \rpro-rich halo stars 
(Truran \etal\ 2002\nocite{tru02}; Cowan \& Sneden 2006\nocite{cow06}; 
Ivans \etal\ 2006\nocite{iva06}) have in general been consistent and 
found to lie in the range $\simeq$~10--15~Gyr.
However, the observed abundance ratio for the star CS~31082-001 (Hill \etal\ 
2002)\nocite{hil02} is significantly higher: Th/Eu~=~0.603 
(log~$\epsilon$(Th/Eu)~=~$-$0.22~$\pm$~0.07), which for our production 
ratio yields a very low age. 
This results from  high U and Th abundances in this star relative to the 
abundances of elements in the range from Ba to the \third\ \rpro\ abundance 
peak, as determined by Hill \etal.
However, the lead abundance -- resulting from the decay of Th and U --  
can provide a strong constraint on the abundance values and insight into the 
synthesis mechanisms of these radioactive actinides.


\subsection{\bf The Actinide/Lead Abundance Ratio in Halo Stars}


Can the apparent single-valued \rpro\ abundance pattern in the 
elements Ba--Pt be expected to extend beyond the Pt--Pb peak? 
A further constraint on the robustness of the \rpro\ in the regime from 
the \third\ \rpro\ peak through the actinides can be provided by observations
of the  Th, U  and Pb abundances in metal-poor stars. 

The actinide chronometers Th, \iso{U}{235}, and \iso{U}{238} 
decay directly into the lead isotopes 
\iso{Pb}{208}, \iso{Pb}{207}, and \iso{Pb}{206}, respectively. 
The lead abundances, therefore, provide a measure of the abundance levels 
of the chronometer nuclei. 
Consider the following quantitative measures. 
The Th/Pb ratio at the time of SS formation 
(Lodders 2003\nocite{lod03}) is (Th/Pb)~=~0.04399/1.4724~= 0.02988. 
The Th/Pb ratio in our \rpro\ calculations is significantly 
higher: (Th/Pb)$_{r-process}$ = 0.095. 
This implies first that the \rpro\ is responsible for of order 30\% 
($\approx$~0.03/0.095) of the SS abundances of the heavy lead 
isotopes \iso{Pb}{206}, \iso{Pb}{207}, and \iso{Pb}{208}.  
The Th and U chronometer abundances in extremely metal-poor \rpro-rich 
halo stars, therefore, also provide a measure of the expected level of 
Pb in these same stars. 
These lead levels, since they are the products of the decays of nuclei Th, 
\iso{U}{235}, and \iso{U}{238}, can in turn be utilized to constrain the ages 
of these stars as well. 
In this context, the high levels of Th and U cited for the halo star 
CS~31082-001 are inconsistent with the abundances of the platinum peak 
isotopes and the limits on the lead abundance for this star (Plez \etal\
2004).\nocite{ple04}
In contrast, the adoption of the upper limit on the Pb abundance for the star 
\bd17\ (Cowan \etal\ 2002) yields log~$\epsilon$(Th/Pb)~$>$~$-$1.48, and thus 
Th/Pb~$>$~0.033 -- this value seems consistent with the abundances of 
the \third\ \rpro-peak elements and the expected \rpro\ production ratio. 
Also consistent are the Th/Pb ratios of 0.024 to 0.042 (depending upon 
adopting one of the very uncertain lead values) that have been reported 
for \cs22\ (Sneden \etal\ 2003). 
More work needs to be done to understand specifically how CS~31082-001  
can have such an overabundance of the actinide chronometers Th and U with 
a correspondingly low Pb abundance. 
More generally it needs to be determined if there are other such stars. 

Abundances of lead in the metal-poor halo stars, while difficult to derive
and ideally requiring space-based observations, offer the promise of more 
refined (and constraining) chronometric age determinations.


\section{CONCLUSIONS}


The rapid \ncap\ process is understood to be responsible for the synthesis 
of approximately half of all of the isotopes present in solar system 
matter in the mass region from approximately zinc through the actinides.
While the general features of this process were identified in the classic
papers by Burbidge \etal\ (1957)\nocite{bur57} and Cameron 
(1957)\nocite{cam57}, our current understanding of the \rpro\ 
remains incomplete.
We have yet to cleanly identify which of the proposed 
astrophysical sites contribute significantly to the observed abundance 
pattern and we have yet to reconcile the apparent duplicity of \rpro\ 
sites with extant models for the operation of the \rpro\ in diverse 
astronomical environments.
In this paper we have explored implications from parameterized waiting-point
synthesis calculations.

Various observations suggest that contributions from two different \rpro\
astronomical sites or environments are   required, for the mass regimes
A~$\lesssim$~130 and A~$\gtrsim$~130 respectively. 
Here,  we have tried to identify the mass number that represents the dividing 
line between these two (``weak'' and ``main'') \rpro\ contributions. 
In this context, our discussions in this 
paper and associated calculations lead to the following conclusions:

\begin{itemize}

\item{} 
The combined elemental and isotopic data on low metallicity 
\rpro-rich stars confirms a robust \rpro\ pattern extending over the 
(A $>$ 130-140) \ncap\ isotopic domain.

\item{} 
We are able to reproduce the total SS \rpro\ abundances with a 
superposition of neutron number densities ranging from log~$n_n$~= 20--28.  
Our calculations indicate that smaller neutron number densities,
(log~$n_n$~= 20--22) that characterize the weak \rpro, are required to 
reproduce the A = 80-100 SS isotopic \rpro\ abundances. 
We estimate that the division between ``weak'' and ``main''
\rpro\ density regimes occurs at log~$n_n$~$\simeq$ 23.0~$\pm$~0.5.
More sophisticated network calculations (Farouqi \etal\ 2006)\nocite{far06} 
appear to bear out our general results.

\item{} The \rpro\ calculations that successfully generate the element 
pattern extending down to A=135 indicate that the production of 
\iso{I}{129} is at a level $\sim$90\% of its 
solar \rpro\ value, relative to the Ba-Pb region. 
Our calculations suggest that the \iso{I}{129}/\iso{Hf}{182} production 
ratio is quite compatible with the anticipated \rpro\ abundance pattern. 
In this context, our results imply that the dividing line (in mass number) 
between the ``weak'' and ``main'' \rpro\ components must necessarily fall 
below \iso{I}{129}. 
We note, however, that observational limitations, so far, prevent, an 
exact identification of the elemental atomic number where the break 
occurs between the ``main'' and ``weak'' \rpro. (Iodine, for
example, has not yet been observed in a  metal-poor halo star.) 

\item{} 
We find that the isotopic fractions of barium, and the elemental Ba/Eu
abundance ratios in \rpro-rich low metallicity stars can only be matched 
by computations in which the neutron densities are in the range 
23~$\lesssim$~log~$n_n$~$\lesssim$~28, values typical of the main \rpro.
For the main \rpro\ our calculations predict a 
solar \rpro\ value of the Ba isotopic ratio, 
$f_{odd}$~$\equiv$ [N(\iso{Ba}{135})+N(\iso{Ba}{137})]/N(Ba) $\simeq$   
0.47--0.50.  
The observed value of $f_{odd}$ = 0.3 $\pm$ 0.21 in one metal-poor halo 
star is consistent with the SS \rpro\ ratio. 
While the uncertainty is large and does not rule out some slight \spro\
production, the elemental ratio of [Ba/Eu] in this star is in agreement
with the SS \rpro\ only ratio.   
Since this star's Ba/Eu elemental abundance ratio is also confirmed by
our calculations, an \rpro-only origin for its Ba isotopes is strongly
indicated.  

\item{} 
In the neutron number density ranges required for production of the
observed solar/stellar \third\ \rpro-peak (A~$\approx$~200),
the predicted abundances of inter-peak element hafnium (Z~=~72,
A~$\approx$~180) follow closely those of \third-peak elements
(osmium through platinum) and lead.
This suggests that abundance comparisons of hafnium to both rare-earth
and \third-peak elements can shed further light on claims of invariance
in the entire heavy end of the \rpro\ abundance pattern.
Hafnium, observable from the ground and close in mass number to the \third\
\rpro-peak elements, could  be utilized as a new nuclear chronometer
pair Th/Hf 
for age determinations, 
particularly for cases where U is not detected. 
In the context of the calculations that reproduce the stable SS 
\rpro\ abundances, we have determined several important chronometric 
production ratios including Th/U, Th/Eu, Th/Pt and Th/\third-peak 
elements and Th/Hf. 
For example, the present computations suggest Th/\iso{U}{238}~$\simeq$~1.5
for this chronometer pair.  
These newly predicted chronometric ratios can then be employed to 
determine ages in stars where Th or U have been detected. 

\end{itemize}


\acknowledgments


We thank Roberto Gallino, Peter M{\"o}ller, Anna Frebel, and Ulli Ott for useful 
discussions, and an anonymous referee for helpful suggestions. 
This work has been supported in part by the Deutsche Forschungsgemeinschaft 
(DFG) under contract KR 806/13-1, and the Helmholtz Gemeinschaft under 
grant VH-VI-061 and 
by GSI (Univ. Mainz F+E-Vertrag (MZ/KLK)).
Support was also provided by the National Science Foundation  
under grants AST 03-07279 (J.J.C.), AST 03-07495 (C.S.), and the Physics
Frontier Center (JINA) PHY 02-16783 (J.W.T.), by the DOE under contract
B523820 to the ASCI Alliances Center for Astrophysical Flashes (J.W.T.),
and at the Argonne National Laboratory, which is operated under contract No.
W-31-109-ENG-38 (J.W.T).


\clearpage
\tablecolumns{7}
\tablewidth{0pt}
\begin{deluxetable}{ccccccc}
\tablecaption{r-Process Isotopic Barium Abundances\label{tab1}}
\tablehead{
\colhead{log n$_n$}                     &
\colhead{\iso{Ba}{134}}                 &
\colhead{\iso{Ba}{135}}                 &
\colhead{\iso{Ba}{136}}                 &
\colhead{\iso{Ba}{137}}                 &
\colhead{\iso{Ba}{138}}                 &
\colhead{$f_{odd}$\tablenotemark{a}}    \\
}
\startdata
20.0 & \nodata & 1.40E-09 & \nodata & 1.71E-11 & 1.32E-10 & 9.15E-001 \\
20.5 & \nodata & 5.96E-08 & \nodata & 4.00E-10 & 3.19E-09 & 9.49E-001 \\
21.0 & \nodata & 3.88E-06 & \nodata & 3.27E-09 & 2.63E-08 & 9.93E-001 \\
21.5 & \nodata & 6.70E-05 & \nodata & 1.62E-07 & 7.90E-07 & 9.88E-001 \\
22.0 & \nodata & 1.04E-04 & \nodata & 1.76E-05 & 9.45E-05 & 5.62E-001 \\
22.5 & \nodata & 8.65E-04 & \nodata & 4.32E-04 & 2.43E-03 & 3.48E-001 \\
23.0 & \nodata & 1.96E-02 & \nodata & 4.31E-03 & 2.45E-02 & 4.94E-001 \\
23.0 & \nodata & 1.96E-02 & \nodata & 4.31E-03 & 2.45E-02 & 4.94E-001 \\
23.5 & \nodata & 4.93E-02 & \nodata & 9.63E-03 & 5.50E-02 & 5.17E-001 \\
24.0 & \nodata & 7.25E-02 & \nodata & 1.54E-02 & 8.64E-02 & 5.04E-001 \\
24.5 & \nodata & 6.35E-02 & \nodata & 2.12E-02 & 1.09E-01 & 4.38E-001 \\
25.0 & \nodata & 2.60E-02 & \nodata & 1.30E-02 & 5.00E-02 & 4.38E-001 \\
25.5 & \nodata & 3.02E-02 & \nodata & 9.39E-03 & 3.32E-02 & 5.44E-001 \\
26.0 & \nodata & 3.80E-02 & \nodata & 1.13E-02 & 5.62E-02 & 4.68E-001 \\
26.5 & \nodata & 2.76E-02 & \nodata & 1.04E-02 & 5.59E-02 & 4.05E-001 \\
27.0 & \nodata & 1.02E-02 & \nodata & 7.63E-03 & 4.16E-02 & 3.00E-001 \\
27.5 & \nodata & 3.42E-03 & \nodata & 3.38E-03 & 1.85E-02 & 2.68E-001 \\
28.0 & \nodata & 2.36E-03 & \nodata & 5.50E-04 & 2.98E-03 & 4.95E-001 \\
28.5 & \nodata & 1.52E-03 & \nodata & 2.90E-04 & 9.01E-04 & 6.68E-001 \\
29.0 & \nodata & 6.99E-04 & \nodata & 2.95E-04 & 9.83E-04 & 5.03E-001 \\
29.5 & \nodata & 5.98E-04 & \nodata & 4.32E-05 & 2.15E-04 & 7.48E-001 \\
30.0 & \nodata & 4.69E-04 & \nodata & 5.14E-05 & 1.67E-04 & 7.56E-001 \\
\enddata

\tablenotetext{a}
{$f_{odd}$ $\equiv$ [N(\iso{Ba}{135})+N(\iso{Ba}{137})]/N(Ba)}

\end{deluxetable}

\clearpage
\begin{deluxetable}{cccccc}
\tablecolumns{6}
\tablewidth{0pt}
\tablecaption{r-Process Production Ratios\label{tab2}}
\tablehead{
\colhead{Ratio\tablenotemark{1}}     &
\colhead{Cowan et al. (1999)}                 &
\multicolumn{2}{c}{Fe-seed}          &
\colhead{Zr-seed}                    &
\colhead{Th(fit2)/X(sol)}            \\
\colhead{}                           &
\colhead{}                           &
\colhead{(fit1)\tablenotemark{2}}    &
\colhead{(fit2)\tablenotemark{3}}    &
\colhead{}
}
\startdata
Th/$^{238}$U           & 1.805 &  1.557 & 1.568 & 1.475  & \nodata \\
$^{235}$U/$^{238}$U    & 1.602 &  1.464 & 1.464 & 1.758  & \nodata \\
Th/Os                  & 0.099 &  0.098 & 0.093 & 0.0735 &  0.0750 \\
Th/Ir                  & 0.092 &  0.095 & 0.089 & 0.0676 &  0.0703 \\
Th/Pt                  & 0.024 &  0.026 & 0.024 & 0.0316 &  0.0360 \\
Th/3$^{\rm rd}$ peak   & 0.016 &  0.017 & 0.0016& 0.0166 &  0.0181 \\
Th/Eu                  & 0.481 &  0.530 & 0.453 & 0.479  &   0.422 \\
Th/Hf\tablenotemark{4} & 0.897 &  0.864  & 0.862 & 0.637  &  0.462\tablenotemark{5} \\
\enddata

\tablenotetext{1}{based upon ETFSI-Q mass model}
\tablenotetext{2}{based upon the least square fit to the
solar system data from A $>$ 83, see text.}
\tablenotetext{3}{based upon the least square fit to the
solar system data from A $>$ 125, see text}
\tablenotetext{4}{based upon the average of 6 different neutron number
densities between 10$^{20}$ and 10$^{30}$}
\tablenotetext{5}{based upon the solar $r$-process value for Hf from
Lawler {\it et al.} (2006b). 
}

\end{deluxetable}

\clearpage

\clearpage

\clearpage

\clearpage

\clearpage

\clearpage

\clearpage
\nocite{sne03,den03,law06a,den06,law06b,den05}

\clearpage

\clearpage

\clearpage

\clearpage

\clearpage

\end{document}